\documentclass[11pt]{article}
\newcommand{\fP}{I\!\!P}
\usepackage{epsfig}
\begin{document}
%\titlepage
%\begin{minipage}[t]{4cm}
\begin{flushleft}
 DESY--00--154\\
 hep-ph/0010345\\
 October 2000
%  hep-ph/00\,12\,xxx
%\end{minipage}
\end{flushleft}
\vspace*{0.5in}
\begin{center}
%\title 
{\Large Diffraction at HERA and the Confinement Problem
%\\[0.5cm]
%\large HERA Measurements  and the Properties of the Strong Force
}\\
%\end{Large}
\vspace*{0.5cm}
%\author
{J. Bartels \\II. Institut f\"ur Theoretische Physik,Universit\"at
Hamburg \\
H. Kowalski \\ Deutsches Elektronen Synchrotron DESY, Hamburg}
\end{center}
%
%\date{  }
%\maketitle

\begin{abstract}
\noindent
We discuss HERA data on the high energy behavior of the total 
$\gamma^*p$ cross section and on
diffraction in deep inelastic scattering. We outline their novelty 
in comparison with diffraction in high energy 
hadron-hadron scattering. As a physical picture, we propose an 
interpretation in terms of QCD radiation at small and large distances: 
a careful study of the transition between the two extremes 
represents a new approach to the QCD confinement problem.    
\end{abstract}

%%%%%%%%%%%%%%%%%%%%%%%%%%%%%%%%%%%%%%%%%%%%%%%%%%%%%%%%%
\section{Introduction}

Quantum Chromodynamics (QCD) is expected to describe the 
strong forces between
hadrons.  In particular the theory should be able to explain high energy 
scattering processes. At short distances, smaller than the 
proton radius, QCD describes the interaction between quarks and gluons 
in an analogous way to that in which  Quantum Electrodynamics  provides 
a quantitative description of the 
interaction between electrons and photons. High energy processes,  
however, are often dominated by the forces at large distances 
(of the order of the proton radius), where a satisfactory 
understanding of QCD still remains a challenge. 
One of the main difficulties is the lack of information about the 
transition from small to large distances. Here results of HERA 
experiments at DESY in Hamburg are providing unexpected help. 

In the past, theoretical efforts to understand the dynamics 
of confinement  
have mainly been concentrated on analyzing the binding forces between 
{\it static} quarks: 
the bound states of a quark and an antiquark or of three 
quarks. The knowledge of these forces, which 
often are conveniently described in terms of an 
 {\it interquark-potential},  would allow the calculation of  
masses and other static properties of hadrons - certainly one of the 
central problems of strong interaction physics.  
However, in {\it high-energy scattering experiments} the 
question of the binding forces shows up in a somewhat different form. 
For example, if two protons collide at high energies, usually 
a large number of tracks of single particles and sometimes of 
jets are seen in the detectors. This is not surprising if one remembers  
the well-known picture of  
sophisticated mechanical Swiss watches: if two such complex objects 
collide head-on at high energies, one certainly expects 
to find, after the collision, only the debris of the watches, and one has 
little hope that the watches remain intact. The higher the energy of the 
collision process, the smaller the probability of finding the incident 
projectiles undamaged after the collision. However, in the 
high energy hadron-hadron scattering processes one also observes a 
rather large fraction of events in which the incoming particles scatter at 
very small angles and remain practically intact 
(called ``diffractive''): they
either  remain completely intact (elastic scattering, about 20 \%) 
or they go into an excited state with the 
same quantum numbers (called ``quasielastic'', about 10 \%) 
Furthermore, these fractions remain almost constant with energy. 
These observations clearly reflect fundamental properties of the binding 
forces inside the hadrons which cannot be described by the concept of the 
interquark-potential. 

In the analysis of a high energy scattering process we should  
distinguish,  as in the static confinement problem, 
between short and long distances. With nucleons as incoming 
projectiles the scattering process is dominated by distances of the 
order of the  proton radius: all our experience with elastic 
scatttering at hadron colliders, therefore, reflects features of QCD 
forces at large distances which we cannot yet calculate.
On the other hand,  HERA experiments  investigate the 
structure of the proton using short wave electromagnetic radiation, 
real and virtual photons at very high energies.
The virtuality of the photon, $Q^2$, varies between 0 and O(10000) GeV$^2$
which means that the photons' sizes can be large or very small.
HERA measurements, therefore, have opened the door to a 
completely new class of reactions which, for the first time, allow the
investigation of both the short and long distance region in 
elastic high energy scattering. 

HERA experiments have shown that the scattering of a 
small photon on the proton is different from what we know
from hadron-hadron scattering. First, the total cross section 
exhibits a considerably stronger rise with energy than observed in 
hadron-hadron scattering. Secondly, the spread of the scattering
system in the transverse direction which can be seen as measure of the
forces keeping the scattering projectiles intact is quite different from hadron-hadron 
scattering. Since HERA allows  the size of the photon to vary,
it is possible to interpolate between short and large distances; whereas in 
the small size region the novel features emerge, in the large size region 
HERA establishes consistency with the basic features of hadron-hadron 
scattering.  
 
Comparisons with QCD calculations show that the small size region
can be described successfully within the framework of perturbation theory.
These calculations of elastic high energy 
scattering at short distances can be formulated as a 
{\it radiation problem}:
the incoming photon creates a quark-antiquark pair which 
radiates gluons and further quark-antiquark pairs. The transition from the 
scattering of a small 
size photon on a proton to hadron-hadron scattering can then be viewed 
as the {\it transition 
of QCD radiation from small to large transverse sizes}.
Thus at HERA we can not only test  our understanding of QCD at short 
distances but also explore the transition into the confinement region.
A careful study of QCD radiation at small and large distances 
presents a new approach to the QCD confinement problem. 

In this article we will review various elastic and quasi-elastic  
processes observed at HERA and discuss their implications for our 
understanding of the QCD forces at small and at large distances. In order 
to illustrate the novelty of HERA results it will be useful to begin 
with a short review of the main properties of ``old'' elastic hadron-hadron
scattering at high energies: they represent the long-distance 
(nonperturbative) part of the QCD forces. We then turn to the 
discussion of HERA results and their interpretation in terms of properties 
of QCD. At the end we briefly comment on the connection with deep inelastic 
structure functions.  
   
\newpage

\section{Properties of hadron-hadron scattering}

We have already mentioned that in hadron-hadron scattering a substantial 
fraction of interactions 
(about 30 \%) is elastic or quasi-elastic, and both the total 
and the elastic cross sections remain almost constant with energy. 
Moreover, certain features of the scattering cross sections have been found
to be {\it universal}, i.e. independent of the species of the incoming hadrons.
The theoretical ansatz  which, within Regge theory, incorporates 
these observations carries the name ``Pomeron''. It was  
invented by the Russian theoretician V.N. Gribov ~\cite{Gribov1} and 
afterwards named after another Russian theoretician Y. Pomeranchuk.
  
For the elastic scattering of  two hadrons $a$ and $b$ at high 
energies and small angles the following economic and convenient 
parameterization of the scattering amplitude  has been found:
\begin{eqnarray}
T_{el}^{ab}(s,t) = i s \beta_a(t) (s/s_0)^{\alpha_{\fP}(t)-1}\beta_b(t)
\label{eq:sast}
\end{eqnarray}
Here $s$ and $t$, in units of GeV$^2$, denote the square of the total 
energy of the hadron-hadron system and the square of the momentum transfer, 
respectively. In particular,
$t=0$ implies  zero scattering angle. 
In our discussion, we restrict 
ourselves to small $t$-values, say $-0.5<t<0$ GeV$^2$. For simplicity we 
have approximated the phase factor by $i$, i.e. we ignore the real part 
of the scattering amplitude. $s_0$ is a hadronic scale which is frequently 
chosen to be of the order of $1$ GeV$^2$. 
The dependence on the species of the incoming hadron is contained in the
form factors, $\beta_{a,b}$, for which it is convenient to use a simple
exponential ansatz. 
The observed energy dependence of the 
cross section can be described by the ansatz for 
 the exponent $\alpha_{\fP}(t)$:
\begin{eqnarray}   
\alpha_{\fP}(t)=\alpha_{\fP}(0) + \alpha_{\fP}'t = 1+\epsilon + 
\alpha_{\fP}' t \, .
\label{eq:alphat}
\end{eqnarray} 
This function has been found to be independent of the species of the 
incoming hadrons $a$ and $b$. The two parameters $\alpha_{\fP}(0)$ 
(or $\epsilon$)
and $\alpha_{\fP}'$, therefore, appear to be fundamental, and they represent 
universal features of the strong forces that describe the binding of 
hadrons (in the jargon the parameters 
$\alpha_{\fP}(0)$ and $\alpha_{\fP}'$ are called  the ``Pomeron intercept''
and ``Pomeron slope'', respectively).
In the analysis of high-energy scattering processes, these two parameters have
a quite different meaning. The first one, $\epsilon $, determines the
energy dependence of the elastic cross section:
\begin{eqnarray}
d\sigma_{el}^{ab}/dt|_{t=0} = \frac{1}{16 \pi} [\beta_a(0) \beta_b(0)]^2 
(s/s_0) ^{2\epsilon}.
\end{eqnarray}
Via the Optical Theorem, 
\begin{eqnarray}
\sigma_{tot}^{ab} = \frac{1}{s} Im T_{el}^{ab}(s,0) 
\end{eqnarray}
it also describes the energy dependence of the total cross section:
\begin{eqnarray} 
\sigma_{tot}^{ab} = \beta_a(0) \beta_b(0) (s/s_0)^{\epsilon}.
\end{eqnarray}
Various experiments have found the value $\epsilon \approx 0.08$ ~\cite{DL}.

\begin{figure}[hbt]
%\vspace{0.5cm}
\hspace{1.0cm}
\epsfig{file=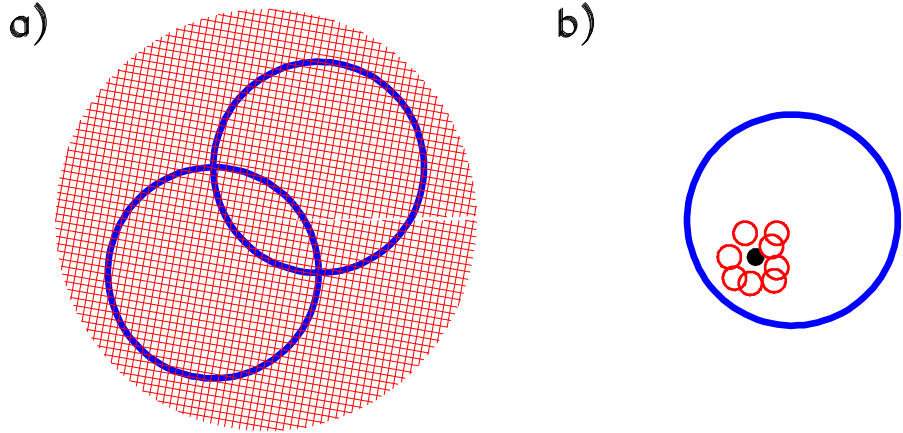,height=7.6cm,width=16cm } 
\caption{\sl (a) Section of the hadron-hadron scattering process in the plane
transverse to the direction of flight. The circles denote the hadrons, the 
shaded area the full interaction region (eq.\ref{eq:bs}): whereas the size of the hadrons
stays fixed, the extension of scattering profile grows with energy; (b) Section of the $\gamma^*p$ scattering process in the transverse 
direction. The big circle denotes the proton, the black dot the virtual 
photon which creates a $q\bar{q}$ pair and then builds up its 
radiation cloud which is denoted by the small open circles.}
\label{fig:disc}
\end{figure}

To understand the meaning of    
the second parameter, $\alpha_{\fP}'$, we first note that 
the ansatz  for the 
scattering amplitude,  eq.~(~\ref{eq:sast}),
 leads to a geometrical interpretation.
 To see this we first write  the elastic cross 
section in the following form  (valid for small $t$-values):
\begin{eqnarray}
d\sigma_{el}^{ab}/dt= \frac{1}{16 \pi} [\beta_a(0) \beta_b(0)]^2 
e^{B(s)t} (s/s_0)^{2\epsilon},
\end{eqnarray} 
where
\begin{eqnarray}
B(s)=2\left( B_{0;a}+B_{0;b}+\alpha_{\fP}' \ln s/s_0 \right)
\label{eq:bs}
\end{eqnarray}
The energy independent terms $B_{0;a}$ and $B_{0;b}$ originate from the 
form factors of the hadrons 
$a$ and $b$; for proton-proton scattering, experimental data require
$B_{0;p} \approx 2-3$ GeV$^{-2}$ (if we identify 
$B_{0;p}= <~r_{em}^2~>~/~6$, the 
often quoted value for the proton radius, 
$R_p=\sqrt{<r_{em}^2>}\approx 4$ GeV$^{-1}$, leads to
$B_{0;p}\approx 2.8$ GeV$^{-2}$). 
$B(s)$ increases with the scattering energy $s$, i.e. the 
exponential decrease of the elastic cross section steepens with increasing
energy (this behavior has been observed experimentally and  is called 
``shrinkage''). 
By introducing the two-dimensional transverse momentum vector 
$\vec{k}$ (with $\vec{k}^2=-t$), the 
impact parameter vector $\vec{b}$, and by writing the scattering amplitude as
a Fourier transform of $f^{ab}(s,\vec{b})$, 
$T_{el}^{ab}(s,t)=4s \int d^2b e^{i\vec{k}\vec{b}} f^{ab}(s,\vec{b})$, we 
arrive at:     
\begin{eqnarray}
f^{ab}(s,\vec{b})= \int \frac{d^2k}{16\pi^2s} e^{-i\vec{k}\vec{b}} 
T_{el}^{ab}(s,t)=
i\frac{\beta_a(0) \beta_b(0)}{8\pi}\frac{(s/s_0)^{\epsilon}}{B(s)} 
e^{- \vec{b}^2/2B(s)}.
\label{eq:sasb}
\end{eqnarray}
From this it follows that $R_{int}^2=<\vec{b}^2>=2B(s)$ describes the
mean-square transverse extension of the scattering system 
(``interaction radius'') or 
the size of the interaction region (Fig.~\ref{fig:disc}a). In proton-proton
scattering, 
experiments find $B \approx 8$ GeV$^{-2}$ at small energies, and it grows up 
to $B \approx 12$ GeV$^{-2}$ at $s\approx 5000$ GeV$^2$~\cite{HadRev}.
This geometrical picture can also be formulated directly for the total cross 
section. From the Optical Theorem it follows that 
\begin{eqnarray}
\sigma_{tot}^{ab}(s) = \frac{1}{s} Im T_{el}^{ab}(s,0) = 
4 \int d^2 b f^{ab}(s,\vec{b}).
\label{eq:stotb}
\end{eqnarray}
The gaussian distribution in eq.(~\ref{eq:sasb})
%$e^{- \vec{b}^2/4B(s)}$, 
suggests a picture in which the scattering system is described as a disc
oriented transverse to the direction of flight with a $b$-dependent opacity:
the mean radius of the disc is proportional to $\sqrt{B(s)}$. 
According to eq.~\ref{eq:bs}, $B(s)$ consists of an energy
independent part which is given by the radii of the projectiles $a$ and $b$,
and a  term which grows with energy: therefore the transverse size of 
the scattering system increases at high energies. This growth is determined 
by the universal parameter $\alpha_{\fP}'$: 
it reflects fundamental properties 
of the strong forces in QCD. Measurements have found the 
value $\alpha_{\fP}' \approx 0.25$ GeV$^{-2}$ ~\cite{DL}.
Following this geometric interpretation the total cross section is 
proportional to the area and to the opacity of this disc.
Because of this optical analogy  
it is customary to name elastic and quasi-elastic 
reactions  ``diffractive processes''.  

Nowadays most physicists have become so  accustomed 
to the observation that the 
hadronic total cross section increases with energy, $\epsilon >0$,
that it seems obvious, because of the Optical Theorem,
to have a substantial amount of elastic or quasi-elastic reactions 
in high-energy processes. 
But one could imagine a different world in which 
the total cross sections fall: this would imply that  the cross sections 
for elastic scattering and diffraction would go to zero too. 
The observation 
that the total cross section slowly increases with energy and that 
diffractive states do not die out at high energies clearly tells us 
something about the 
binding forces which keep quarks together: they act in 
such a way that hadrons often succeed in surviving 
the collision and stay intact.
The second parameter, $\alpha_{\fP}'$, which has the dimension of the square 
of a length and describes the growth of the transverse extension, 
$B(s)$, of the scattering system, reflects the strength of the binding 
forces at high energies. This parameter characterizes the confinement 
forces in QCD as they manifest themselves in elastic high-energy scattering.

Before the advent of QCD theoreticians have already tried to develop an 
intuitive understanding of these observations 
~\cite{Bjorken,Feynman,Gribov2}. These attempts resulted 
in a physical picture for the space-time evolution of 
an elastic high-energy scattering process (via the Optical Theorem it then 
also leads to an intuitive understanding of the total cross section). 
Using the language of QCD, this (only qualitative) picture can be 
summarized in the following way. The two hadrons 
with equal but opposite momenta which 
undergo an elastic scattering process at 
high energies break up into clouds of radiation. Within QCD
one can imagine that these clouds consist of soft gluons and quark pairs 
radiated in cascade-like processes; each gluon carries away 
a small fraction of 
the large momentum of the incoming hadron. These ``nonperturbative'' 
gluons are not small (in the transverse direction), 
and they cannot be identified 
with the partons seen in short distance processes.
The formation of this radiation cloud has started long before the hadrons 
enter the collision process: the higher the momenta of the colliding 
particles the more 
time is available for the formation of the radiation clouds. 
As suggested by field theoretical investigations, the interaction between 
the hadrons is mediated by the interaction between 
those parts of the radiation clouds which carry the 
smallest momentum fractions. To emphasize their importance Richard Feynman  
introduced the notion of ``wee partons''. The 
distribution of these wee partons (both in transverse space and in momentum) 
plays the central role in the scattering process. 
This parton-like picture of the scattering process leads naturally to the 
interpretation that the growth of the transverse size of the interaction 
region, $B(s)$, with increasing energy is the result of a ``diffusion'' 
process (indicated by the shaded area in Fig.~\ref{fig:disc}a). 
The wee partons (which at large transverse distances are better described
as a ``pion cloud'') 
diffuse towards the surface of the interaction region with
the mean square free path length $\alpha_{\fP}'$. The variable  
$\ln s$ plays here the role of time. From what has been said 
before one should expect that a complete understanding of 
these wee partons comes close to 
a full solution of the complex QCD dynamics: the reassembling of the wee 
partons after the collision process and the subsequent formation of the 
outgoing bound state, identical to the incoming hadronic states, 
can be understood only if we understand the strong binding forces. 
The fact that the two parameters mentioned 
before are universal strongly indicates  that the wee partons inside the 
radiation cloud are a fundamental aspect of the strong forces in QCD.

Until now it has not been possible to make this picture of a hadronic 
high-energy scattering process more precise.
At the moment we do not know of any possibility to  
calculate the two fundamental parameters, 
$\alpha_{\fP}(0)$ and $\alpha_{\fP}'$,
 from ``first principles''. This is 
not surprising if we remember that the 
language of quarks and gluons 
(in short: perturbative QCD) can be applied only at small distances
and if 
we recapitulate the geometry of the 
elastic high-energy scattering process.  From the radiation picture that we 
have introduced it follows that longitudinal and transverse 
degrees of freedom play quite different roles. Since the formation of the 
radiation clouds start a long time before the collision 
(and far away from the interaction region), the longitudinal 
distances cannot be small. Moreover, the longitudinal extension grows
with increasing energy.  
Theoretical investigations, on the other hand, indicate that it 
is the size of the transverse distances which determines 
the magnitude of the strong coupling constant and thus the applicability of 
perturbative QCD. Incoming hadrons
- even before the beginning of any radiation process - 
have transverse extensions which are large: 
for distances of the order of a 
proton radius the strong coupling constant is not small and 
perturbative QCD does not apply. On top of the ``static'' hadronic 
radii, the radiation and formation 
of wee partons lead to a further (energy dependent) increase in the 
transverse direction and make the situation even worse. 
It may help to visualize the elastic hadron-hadron forward-scattering 
process as taking place inside a long cylindrical tube with its 
axis along the direction of flight. In hadron-hadron scattering this tube
has a large diameter. 
In a Gedanken-experiment one might imagine having two  
incoming particles of small transverse sizes: 
for not too high energies, their scattering would be 
confined to a tube with a very small diameter (a ``femto-tube''), and we 
could hope to be able to calculate the radiation cloud and the scattering 
cross section within perturbative QCD. 

It was a surprise that HERA measurements can be viewed as a 
(partial) realization of this Gedanken-experiment. For most readers, 
HERA is 
the machine which measures deep inelastic structure functions of the proton.
It has turned out that these measurements - when interpreted slightly   
differently - have opened a new door to the understanding of the strong 
interactions. In the following section we shall discuss how the 
short-distance  results obtained from 
HERA measurements differ from the large-distance  
dominated scattering processes seen in hadron-hadron colliders. 
Our discussion will be mainly qualitative; a few theoretical remarks will be 
made in section 3.4.      

\section{Properties of the  $\gamma^* p$ processes  at HERA}

Let us first recall that at  HERA the electrons collide with protons 
by emitting highly energetic 
photons which then hit the proton. Hence, by studying electron-proton 
collisions, in fact, we  
investigate photon-proton scattering processes at high energies (in short: 
$\gamma^*p$) where the photon is  highly virtual. The energies 
of these photons, in the proton rest system, can go up to 
50000 GeV. The new feature at HERA  
is the virtual photon which, compared to $pp$ scattering, has replaced one of 
the incoming hadrons and 
which leads to a remarkable advantage. The photons emitted by the electrons 
can have very small transverse sizes, 
and HERA offers the possibility to control it.
 The possible transverse 
extensions range from a proton size down to about a tenth or a 
hundredth of it.
Such a small photon hits the proton like a sharp ``needle''; this is to be 
compared with a pp-collision, where two ``fat'' and complex projectiles
collide with each other. In this way HERA has brought us closer to the 
Gedanken-experiment described above: one of the 
incoming scattering projectiles, the virtual photon, has a small  
transverse extension. 
Following our discussion above, we should be able to measure (and 
compare with our theoretical understanding) the formation of the 
radiation cloud, the emission of gluons and quark-antiquark pairs, at least 
in the immediate vicinity of the small photon.

\subsection{The total  $\gamma^* p$ cross section}

\begin{figure}[p]
%\vspace{0.5cm}
\hspace{2.5cm}
\epsfig{file=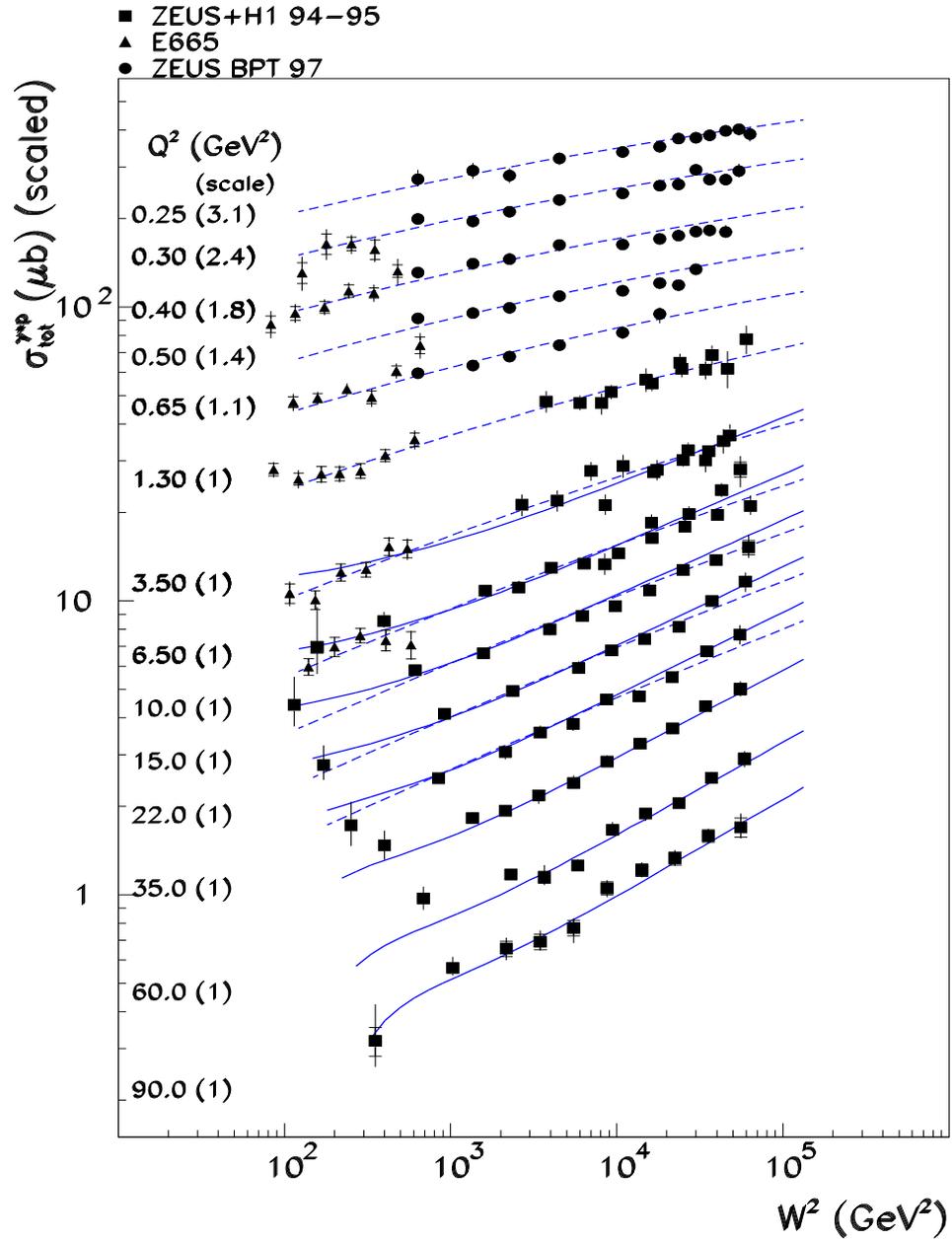,height=18cm,width=14cm} 
\caption{\sl $\gamma^* p$ cross section as a function of $W^2$ at various
$Q^2$. The values of $Q^2$  are shown on the left side  
together with the scale factor applied
to the data for a better visibility. The full line shows a QCD-fit ~\cite{MRS}, the dashed line shows 
a fit by a model ~\cite{GBW}.}
\label{fig:sigtotw2}
\end{figure}
\begin{figure}[hbt]
%\vspace{0.5cm}
\hspace{2.5cm}
\epsfig{file=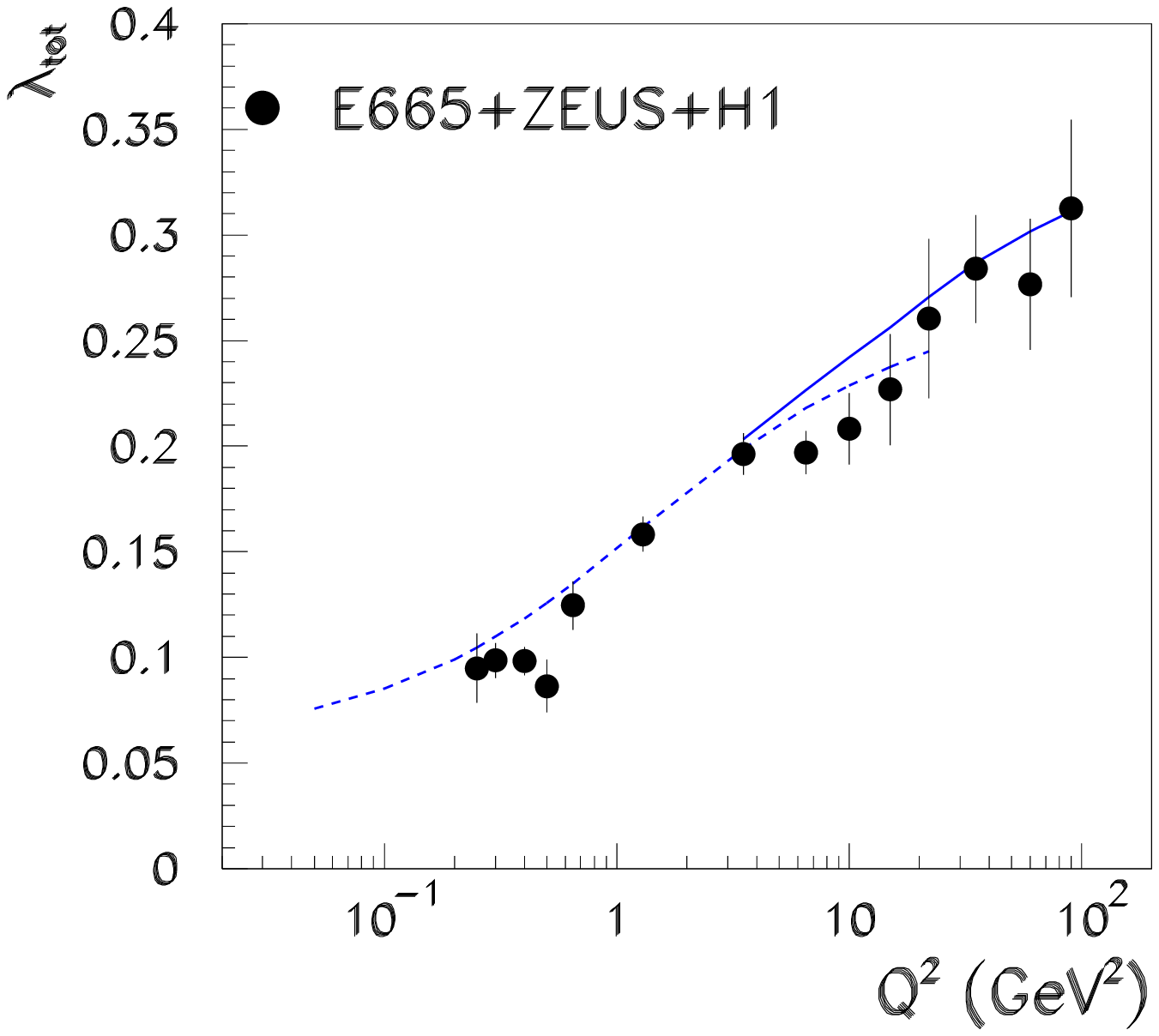,height=10cm,width=10cm} 
\caption{\sl The exponent $\lambda_{tot}$ in the parameteriztion 
$\sigma_{tot}^{\gamma^* p} \sim (W^2)^{\lambda_{tot}}$, plotted as a function 
of $Q^2$.The full line shows a QCD-fit ~\cite{MRS}, the dashed line shows 
a fit by a model ~\cite{GBW}.}
\label{fig:lam}
\end{figure}

One of the most important
observations at HERA is the measurement of the total $\gamma^* p$ 
cross section as a function of the photon virtuality, 
$Q^2$, and of the energy of the $\gamma^* p$ system, 
$W$ (in deep inelastic 
scattering it is customary to denote this energy by $W$ instead of 
$\sqrt s \,$). 
Results are shown in Fig.~\ref{fig:sigtotw2}~\cite{ref.sigtot}.
The variable $\sqrt{ Q^2}$ determines the transverse 
distance inside the proton that the photon can ``resolve'',
$d\approx \frac{2 \cdot 10^{-14}cm}{Q (GeV)}$.  
Beginning at the top of our plot 
with small $Q^2$-values (large transverse sizes) the 
photon resembles a hadron, e.g. a $\rho$-meson. 
Further below, with increasing $Q^2$-values, the photon 
shrinks and becomes more and more point-like. Following  the rise 
of the total cross section in $W$ as a function of $Q^2$, we observe a 
striking 
change. At small $Q^2$ (our plot begins at $Q^2 = 0.25$ GeV$^2$) 
the behavior is still very much the same as in hadron-hadron scattering 
processes. With increasing $Q^2$, the rise in $W$ becomes stronger 
(at the same time the overall magnitude of the cross section decreases). If we 
parameterize the $W$-dependence by a power law, $\sigma_{tot}^{\gamma^*p}
\sim (W^2)^{\lambda_{tot}}$, 
this behavior is translated into the $Q^2$ dependence of the exponent 
$\lambda_{tot}$ (Fig.~\ref{fig:lam}). 
For small $Q^2$ we find ourselves close to 
the hadronic value $\lambda_{tot} = \epsilon = 0.08$, whereas for larger 
values of $Q^2$ the exponent $\lambda_{tot}$ increases substantially.
In this plot the l.h.s. belongs to the hadronic world, the r.h.s. to the 
small-distance world where pQCD is expected to apply.

QCD calculations allow  the results of the small-distance 
part (large $Q^2$ values) to be interpreted in 
terms of QCD radiation. In particular, 
the observed growth of the total cross section with energy can be explained 
in terms of gluon radiation which can be visualized in space and time.
We find it advantageous to view the 
$\gamma^* p$ interactions in the 
proton rest frame rather than in the more usual infinite 
momentum frame used 
in the partonic interpretation of deep inelastic scattering,
in which the proton is fast (see Section 4). 
The measured and 
computed cross sections are, of course, independent of the coordinate 
system; however, the picture of the process and its intuitive ease of
understanding change, depending  on the choice of the system. 
The space-time picture of gluon radiation will be 
formulated for the elastic $\gamma^*$-proton scattering amplitude;
the total $\gamma^*$-proton cross section follows then via the Optical 
Theorem. 

\begin{figure}
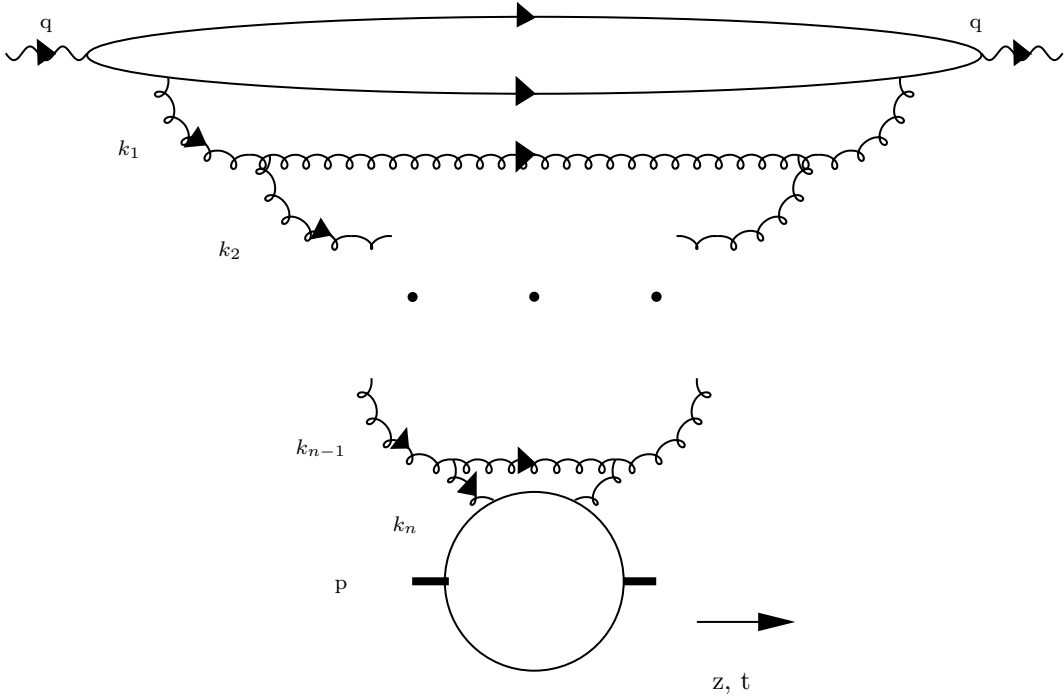

\begin{center}
\input kasten1.pstex_t
\end{center}
\caption{\sl Space-Time diagram of the elastic $\gamma^* p$ scattering
process in the proton rest frame. }
\label{fig:space-time}
\end{figure}

Viewed in the proton rest system, the incoming photon first fluctuates into  
a quark-antiquark pair. Subsequently, the quarks  start to radiate 
gluons. The transverse size of the $q\bar q$ pair depends on the 
photon virtuality, $Q^2$: for our discussion we will be interested in 
fluctuations of the size $\vec{r}^2 \approx 4/Q^2$ (or slightly larger). 
%Fluctuations with much larger sizes are important, too, but they contribute 
%considerably less to the increase with energy. 
For $Q^2 > O(10)$ GeV$^2$, such a $q\bar q$ pair is    
considerably smaller than the proton size of $R_p \approx 4$  GeV$^{-1}$.
For such small distances the strong coupling constant is small, and 
perturbative QCD can be used reliably.
Rather than describing the standard QCD calculations which are done in 
momentum space, we find it instructive to translate them into 
space and time variables. 
The scattering process is illustrated in 
Figure~\ref{fig:space-time}. The photon moves from the left to the right,
and the interaction with the proton takes place at time zero at $z \approx 0$. 
The lifetime $\tau$ of the $q\bar q$ fluctuation is given by (see Section 3.4):
\begin{eqnarray}
\tau = 1/m_p x.
\end{eqnarray}
The value of $\tau$ becomes large for small $x$:
$x$ is connected with the energy
of the $\gamma^* p$ scattering process by $W^2/Q^2 = 1/x -1$, i.e. 
small $x$ values mean large energies (in a reference frame where the proton carries a large 
momentum the variable $x$ denotes the momentum fraction 
with respect to the proton momentum of the quark struck by the photon).   
At HERA, for virtual photons with $Q^2 =10$ GeV$^2$, the $x$ values range 
between $10^{-2}$ and $10^{-4}$. The $q\bar q$ pair is created at the time $-\tau /2$ before the 
interaction with the proton, which corresponds to
the almost macroscopic distance of up to 1000 Fermi 
away from the interaction point. 
Pictorially speaking this means that the $q\bar q$ pair exists 
for a long time 
and travels a substantial distance while radiating gluons before it finally 
annihilates back into the final-state photon. 

The description of the subsequent radiation processes follows from the 
analysis of perturbative QCD: the leading QCD diagrams, 
when translated into 
time-ordered (old-fashioned) perturbation theory (for  more details see
Section 3.4), lead to a cascade-like time-ordered emission process shown
in Fig.~\ref{fig:space-time}. Some time after the
creation of the $q\bar{q}$ pair, the first gluon with momentum $k_1$ is 
emitted, which subsequently emits the second gluon $k_2$ etc. 
Whereas the first gluon still 
has a small transverse size and carries a sizeable
fraction of the  photon's longitudinal momentum, the later gluons have 
larger transverse extensions and carry smaller momentum fractions.
By the time
the $q\bar{q}$ pair reaches the proton, it is surrounded by a system of
gluons and has built up its ``radiation cloud''. 
 The perturbative picture of the cascade 
processes breaks down when the transverse momentum of the
last gluon becomes too small. The precise value, $Q_0^2$, at which  
 the perturbative development of the cascade stops, is not known. 
It is customary to assume for 
$Q_0^2$ a value of the order of a few GeV$^2$, i.e. the corresponding 
transverse distance is smaller than the proton radius. 
The interaction with the proton at rest is through the last
gluon with smallest fraction of the photon momentum and largest transverse 
size: its interaction with the proton cannot be described in perturbation 
theory. Most likely the last gluon further evolves 
 into a nonperturbative cascade of wee partons. After the 
interaction with the proton it recollects the emitted gluons.
The final $q\bar{q}$ pair annihilates into the outgoing photon at time 
$\tau/2$. 

The most striking consequence is that this perturbative 
gluon radiation explains the observed rise of the total cross section. 
This rise, therefore, is a measure of the intensity of gluon radiation,
and it only weakly depends upon the nonperturbative interaction of the last
gluon with the proton. 
The theoretical analysis of these emission processes results in a 
simple ``QCD radiation formula'' (see section 3.4), 
which leads to a growth of 
the total cross section with energy. 
There is a simple way to illustrate this property:
assuming, for simplicity, that in the cascade of gluon emissions all 
transverse momenta are of the same order, and that there is no energy 
dependence 
coming from the interaction of the last gluon with the proton, one easily 
obtains the power law:
\begin{eqnarray}
\sigma_{tot}^{\gamma^*p} \sim \left( W^2 \right)^{\lambda}. 
\label{eq:xtol1}
\end{eqnarray}
Although the real  calculations are slightly more complicated, 
the simple power law in eq.(~\ref{eq:xtol1}) describes the data very well, 
and QCD predicts the right order of magnitude for the exponent $\lambda$. 
Precise QCD descriptions of HERA data are based upon the radiation
formula (see eq.(~\ref{eq:sigtot}) in Section 3.4), with
higher order corrections  taken into account.
An important example of these corrections is the creation 
of quark-antiquark pairs inside the radiation cloud: one 
of the horizontal gluons in Fig.~\ref{fig:space-time} could be replaced by
such a quark pair. Moreover, in a precise QCD description of HERA results, 
data-motivated assumptions are made for the nonperturbative 
interaction of the 
last gluon with the proton. The results of such a computation for
$Q^2>3.5$ GeV$^2$, shown by the full lines in 
Fig.~\ref{fig:sigtotw2} and Fig.~\ref{fig:lam}, 
give a good description of the data. This success of perturbative QCD demonstrates    
that, for the energy dependence of $\gamma^*p$ scattering, perturbative QCD 
radiation leads to a much stronger 
growth of the cross section with energy than the nonperturbative 
interactions of the last gluon with the proton 
(inside the circle of Fig.~\ref{fig:space-time}). 

How could this picture of the gluon radiation around the $q \bar q$ pair
match with the wee parton picture outlined in the previous section?
This is    
certainly one of the main theoretical challenges raised by 
the HERA experiments; at present we can only formulate first ideas and 
conjectures.
In order to safely apply perturbative QCD it was essential that the 
initial photon carried a large virtuality 
$Q^2$, and the transverse extension of the produced $q \bar q$ pair was small. With each step of gluon 
emissions, the transverse size of the radiation cloud gets a bit larger, 
and when it reaches the scale $\sim 1/\sqrt{Q_0^2}$, we enter the 
nonperturbative region (in Fig.~\ref{fig:space-time} marked by the circle). 
This suggests that the hadronic-size wee partons which become relevant for 
the scattering of the last gluon with the proton 
can be viewed as a continuation of the perturbative gluon 
cloud into the region of larger transverse distances: the wee partons 
(which have larger transverse sizes than the perturbative gluons) are 
certainly present and abundant in this interaction. But their  
contribution to the rise of the cross section with $W$, at large $Q^2$,
is considerably weaker than  of the perturbative partons.   
It seems worthwhile to stress that the perturbative QCD radiation cloud 
supports some of the essential features of 
the hadronic wee parton picture 
which, so far,
has remained qualitative: the long time scale of the formation of the
radiation cloud, and the fact that it is the gluon with the smallest 
longitudinal momentum fraction which enters the interaction with the 
proton.  

HERA data on $\sigma_{tot}^{\gamma^*p}$ have shown that, 
when $Q^2$ changes from large to small values, 
one observes a fairly smooth 
transition from the small-size process with the perturbative radiation 
cloud to the hadron-hadron-like scattering process governed by the 
nonperturbative wee partons. It therefore seems natural to search, as a 
function of $Q^2$, for a continous transition of the perturbative 
radiation formula to the nonperturbative wee partons . 
When $Q^2$ decreases, the following substantial changes are expected. 
With increasing size of the $q\bar{q}$ pair the perturbative cascade 
becomes shorter, i.e. it takes
fewer steps of emission before the transverse sizes of the radiated gluons 
reach the boundary of the short-distance region. Furthermore, it becomes 
likely that the perturbative radiation cloud develops a second cascade.  
As an example, one of the horizontal gluons in Fig.~\ref{fig:space-time} 
might start its own cascade, in the same way as the $q\bar{q}$ pair did. 
In such a scenario the perturbative radiation cloud becomes denser, 
and a novel state of high-density partons is formed. 
The creation of this state 
can then be shown to lead  to a weakening of  
the rise in energy, and this slowdown  takes place already inside the short 
distance part of the radiation cloud, i.e. before long distance effects come 
into play. This could explain the observed flattening of 
$\sigma_{tot}^{\gamma^*p}$ with decreasing $Q^2$.
As a consequence, the short-distance part of the radiation cloud looses its
dominance over the large size wee partons, and   
the wee partons begin to play the role known from hadron-hadron scattering.
Thus the $\gamma^* p$ process at low $Q^2$ shows a similar  
 energy dependence as observed in  hadron-hadron scattering.

\subsection{Diffractive processes at HERA}       

The analysis of the total cross section $\sigma_{tot}^{\gamma^*p}$ has 
established that QCD calculations correctly describe the observed energy 
growth and lead to the picture of QCD radiation at small transverse 
distances. More has been 
learned by investigating  diffractive final states in $\gamma^* p$ 
scattering. In the following we will first 
illustrate that the measurement of the energy dependence of diffractive cross 
sections fully confirms and extends the interpretation of the total cross 
section in terms of the radiation cloud. 
We then turn to the measurement of the transverse interaction sizes   
which cannot be deduced from the total cross section alone: 
this is where diffraction at HERA has become most essential.

In the context of deep inelastic electron-proton scattering, 
diffraction denotes final states where the quark-antiquark pair created 
by the photon remains well-separated from the proton 
(the quark pair can also 
be accompanied by one or more radiated gluons), and turns into a final state 
which has the same 
quantum numbers as the initial virtual photon.  Examples of 
diffractive final states at HERA include vector mesons or pairs of jets. 
As to the proton, in 
diffraction  it either stays intact or turns into an excited 
state with identical quantum numbers. As a typical feature of the 
diffractive final state, the invariant mass of the hadronic final state 
of the photon,
$M_X$, is substantially smaller than the total energy of the      
$\gamma^*$-proton system. In the detector this leads to a visible gap  
between the diffractive final states of the photon and of the proton 
(``rapidity gap''). Before HERA it was not expected that diffractive final 
states in deep inelastic scattering would play a significant role: the 
energetic virtual photon smashes the proton into pieces, the final state 
contains large numbers of particles or jets, and it was expected that the 
$q\bar{q}$-pair created by the photon could hardly be separated from the 
rest of the final state.
In contrast to this expectation, a sizable fraction of diffractive 
events was observed.

\begin{figure*}[p]
\centerline{
\epsfig{file=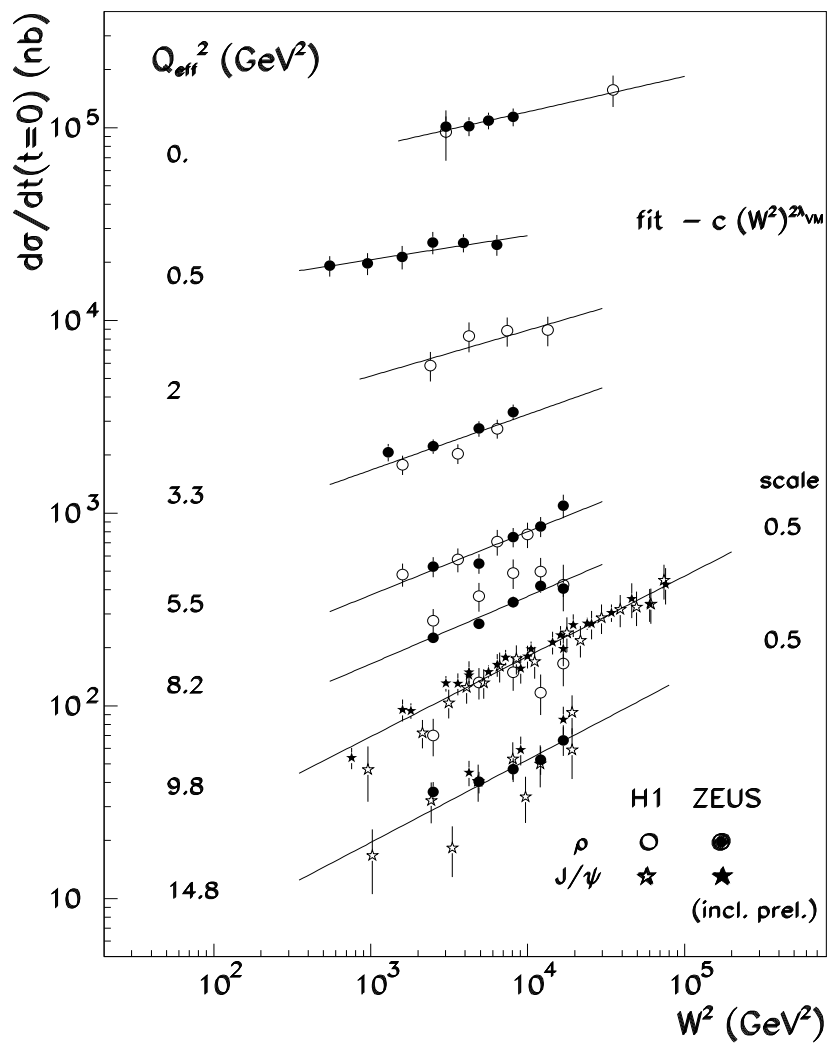,height=16cm,width=12cm}
}
\caption{\sl  Cross sections for diffractive   $\rho$  and  
$J/\Psi$ production in the $\gamma^* p$ processes. The data are plotted
at a scale $Q^2_{eff}$, shown on the left hand side of the plot and defined 
in the text. On the right hand side the scale factor is shown which was 
applied to the data at $Q^2_{eff} = 9.8$ and $14.8$ GeV$^2$ for better visibility.}
\label{fig:zeus.vmbw}
\end{figure*}

We start our discussion of diffractive processes with vector meson
production, $\gamma^* p \rightarrow Vp$. At HERA the production of 
various species of vector mesons has been observed. We will concentrate our discussion on the 
cross sections for the $\rho $ and $J/\Psi$ production. 
Similarly to the elastic process, $\gamma^* p \rightarrow \gamma^* p$ in 
Fig.~\ref{fig:space-time}, the initial photon first creates a 
$q\bar{q}$-pair which soon surrounds itself by a radiation cloud.
The gluon cloud interacts with the proton in the same way as in the elastic 
process. However, after the interaction with the proton and the 
re-collection of the emitted gluons an important difference comes into play:
rather than annihilating into a point-like virtual photon (as in 
Fig.~\ref{fig:space-time}), the $q\bar{q}$-pair now creates a vector meson
which can have a small or a large transverse size. 
The size $r_V$ of a vector meson is determined by its quark content, 
$r_V \sim 1/m_q$ (where $m_q$ denotes the quark mass), and by its wave 
function. Because of the heavy charm quark, $J/\Psi$ production belongs
to the class of small size $q\bar{q}$-pairs (even for the case where the 
initial photon has a small $Q^2$ value). The $\rho$ meson, on the other hand,
is composed of light quarks, and the size of the $q\bar{q}$ is large; 
only a point-like photon (with large $Q^2$, 
and with a particular polarization) 
will help to decrease the transverse size of the $q\bar{q}$ pair. 
HERA results for the energy dependence of vector meson production 
cross sections are shown in   Fig.~\ref{fig:zeus.vmbw}~\cite{ref.vm}. 

\begin{figure*}[htb]
\centerline{
\epsfig{file=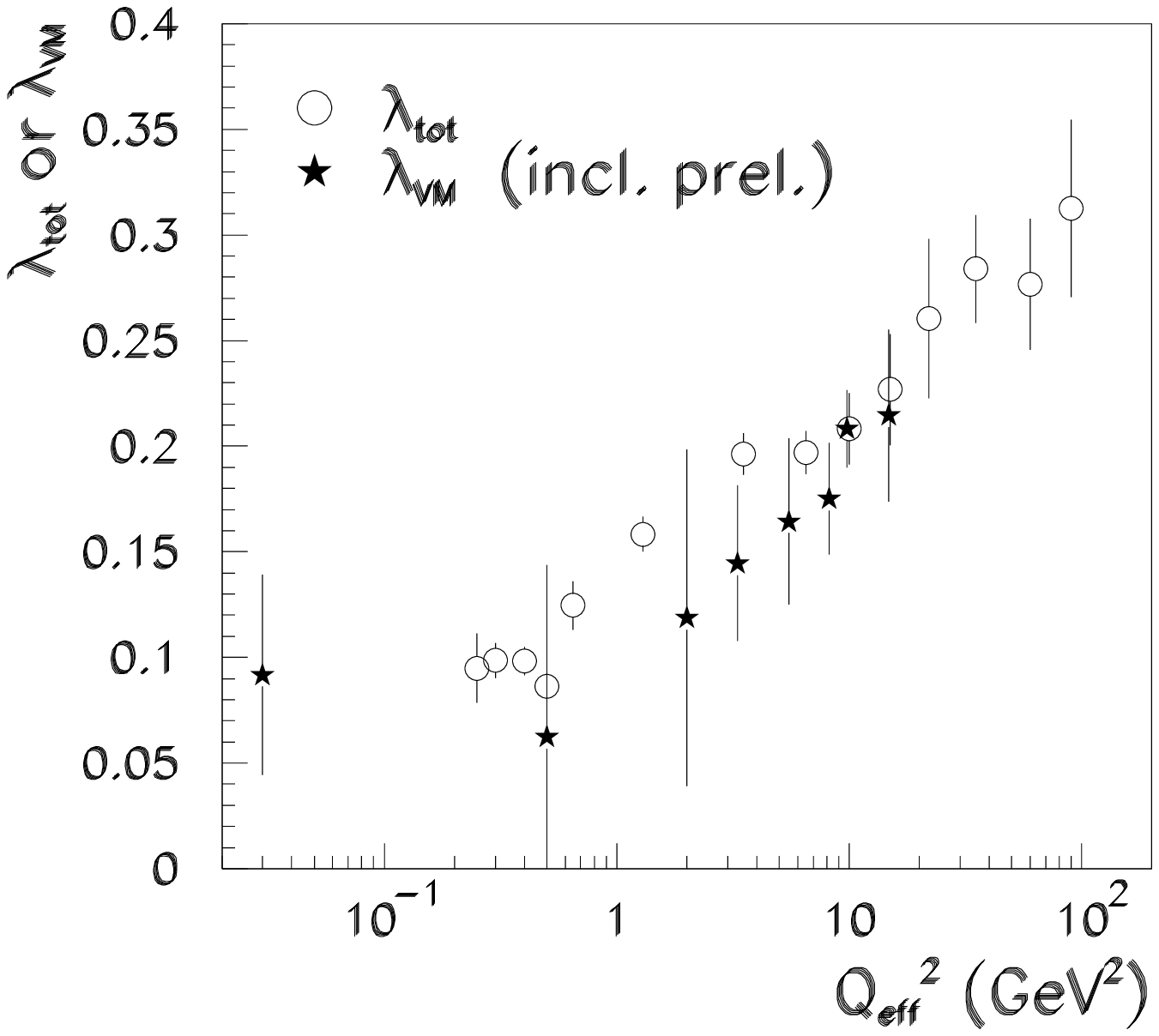,height=10cm,width=10cm}
}
\caption{\sl Comparison of the exponents $\lambda_{VM}$ and $\lambda_{tot}$ 
as a function of the effective $Q^2_{eff}$;  $Q^2_{eff}= Q^2$ for the 
$\rho$ production and  $Q^2_{eff}= Q^2+M^2_{J/\Psi}$ for the 
$J/\Psi$ production. The exponents 
$\lambda_{VM}$ and $\lambda_{tot}$
characterize the 
growth with energy of the diffractive vector meson and total cross section.}
\label{fig:lamvm}
\end{figure*}

For the light vector meson
($\rho$) we have, for $Q^2=0$, a weak energy growth of hadronic cross 
sections. With increasing $Q^2$, the energy dependence becomes stronger.
For heavy quarks ($J/\Psi$) the cross section also exhibits the strong
increase of a small-size system. In order to make this behavior more
quantitative, we parameterize the cross sections as   
\begin{eqnarray}
d\sigma_{VM}^{\gamma^*p}/dt \sim (W^2)^{2\lambda_{VM}}.
\label{eq:xto2l}
\end{eqnarray}
In both cases (light and heavy vector mesons) the exponent $\lambda_{VM}$ 
grows with decreasing size of the
photon (i.e. with increasing $Q^2$). Because of the heavy charm quark mass,
the production of $J/\Psi$ at $Q^2=0$ corresponds to $\rho$-production  
at a higher $Q^2$-scale. A quantitative comparison can be made if  
we introduce suitable momentum scales. 
An appropriate  choice of scale for the $J/\Psi$ production is 
$Q^2_{eff} =Q^2 + M_{J/\Psi}^2$, 
whereas for the $\rho$ production we simply use 
$Q^2_{eff}= Q^2$. Figure ~\ref{fig:zeus.vmbw} shows that 
the cross sections for $\rho$ and $J/\Psi$ production are almost the 
same if we use the  scale $Q^2_{eff}$.
Figure ~\ref{fig:lamvm} shows the values 
of $\lambda_{VM}$ determined from 
data of Fig.~\ref{fig:zeus.vmbw}, at fixed values of $Q^2_{eff}$,
and compares them to the $\lambda_{tot}$ values of Fig.~\ref{fig:lam}. 
The figure shows that the rise of the cross sections 
in different  photon-proton processes 
is strikingly similar in the
whole observed $Q^2$ region. We take this similarity as  
direct experimental 
evidence for the universality of the radiation cloud, not only in the high 
$Q^2$ region but also in the transition to the low $Q^2$ region.

Another important observation at HERA is the measurement of the 
inclusive diffractive cross section, i.e. of the sum of cross 
sections for all processes which exhibit a rapidity gap.  
\begin{figure*}[tbh]
\centerline{
\epsfig{file=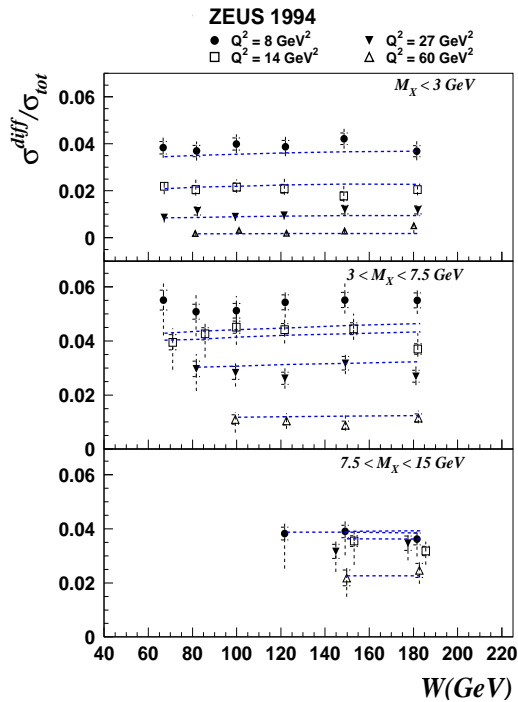,height=10cm,width=8cm}}
\caption{\sl Ratio of diffractive and total cross sections at fixed values
of $Q^2$, for different regions of the invariant diffractive mass $M_X$. 
The lines belong to a model ~\cite{GBW} described in the text.}
\label{fig:zeus.rdifftot}
\end{figure*} 
In this measurement the sum extends over very different diffractive final 
states,
vector mesons, open $q\bar{q}$-pairs which turn into jets, 
$q\bar{q}g$ systems, and others. In order to analyze the space-time evolution 
of these processes we once more 
return to our picture of the elastic $\gamma^*p\to\gamma^*p$ process
(Fig.~\ref{fig:space-time}) and perform suitable substitutions of the
final state photon.
All these reactions have in common that the initial photon creates
a $q\bar{q}$-pair which soon surrounds itself by a radiation cloud. In 
course of the interaction with the proton at rest, however, an important 
difference emerges: in the elastic case and in the case of diffractive
vector meson production, the transverse size of the $q\bar{q}$ pair was 
strongly influenced by the outgoing diffractive system 
(virtual photon or vector meson). For the inclusive cross section, on the 
other hand, the diffractive final state is free to choose its 
transverse size, i.e. it can either be small or large. 
Examples of small-size systems are, once more, the diffractively produced 
vector particles with heavy quarks.
An example of a large-size system is the diffractive  production of 
two  jets of light quarks, which are
aligned along the $\gamma^*p$ scattering axis.
If we follow our discussion above, we expect, for the former case, the 
cross section to exhibit a strong energy increase, $\sigma_{diff} \sim
(W^2)^{2\lambda_{tot}}$.
In the latter case, however, we expect the process to be 
rather hadron-like, with a slowly-rising cross section, $\sigma_{diff} \sim
(W^2)^{2\epsilon}$. The inclusive diffractive cross section sums over 
all these diffractive states with different sizes. Hence, if we define
the exponent $\lambda_{diff}$ by $\sigma_{diff} \sim
(W^2)^{2\lambda_{diff}}$,  the energy dependence 
should be somewhere in between the two extremes:              
$$\epsilon < \lambda_{diff} < \lambda_{tot}.$$
Correspondingly, the mean transverse extension  of the inclusive 
diffractive final state 
lies somewhere between small and large sizes. Before HERA it was generally 
expected that the large-size configurations would give by far the dominant 
contribution, and $\lambda_{diff}$ would be very close to the value 
$\epsilon$. In contrast, measurements at 
HERA (Fig.~\ref{fig:zeus.rdifftot},~\cite{ref.rdiff}) 
show the striking result that, 
at fixed $Q^2$, the ratio of the inclusive diffractive cross 
section and the total $\gamma^*p$ cross section is nearly constant,
i.e. $\lambda_{diff} \approx \lambda_{tot}/2$. Since this value 
is significantly larger than 
$\epsilon$, small-size contributions
clearly are not negligible. Whereas the above inequality 
is a rather straightforward consequence of merely qualitative arguments,
an explanation of Fig.~\ref{fig:zeus.rdifftot} has come from an interesting 
theoretical development which will be discussed in section 3.4.

\subsection{Transverse Sizes}

In our discussion of HERA results so far, we have 
limited ourselves to the forward direction $t=0$. The mean transverse
size $\vec{r}$ of the $q\bar{q}$-pair has played the important role of
discriminating between ``small-size'' and ``large-size'' projectiles.
But in our brief summary of hadron-hadron scattering, we discussed another
transverse scale, the impact parameter $\vec{b}$ with its mean square value
being proportional to $B(s)$: it describes the transverse extension 
(interaction radius) of the scattering system of hadron $a$ and $b$ 
(Fig.~\ref{fig:disc}a). 
At HERA the analogue of this parameter is the transverse extension 
of the scattering system consisting of the quark-antiquark pair and the 
proton at rest (Fig.~\ref{fig:disc}b).
To access this parameter experimentally, we need to measure nonzero 
scattering angles, i.e. the cross sections at $t\neq0$. 
This region cannot be reached for 
the elastic $\gamma^*p$ scattering process.              
Because of the Optical Theorem, the measurement of the total cross section
 provides only the scattering amplitude at $t=0$. Here diffractive final 
states exhibit the unique feature of permitting the measurement of the $t$ 
dependence. 
In particular, in  elastic vector meson production, $\gamma^* p\to Vp$,
the transverse momentum  of the outgoing particles
with respect to the $\gamma^* p$ axis can be directly measured.

\begin{figure*}[htb]
\centerline{
\epsfig{file=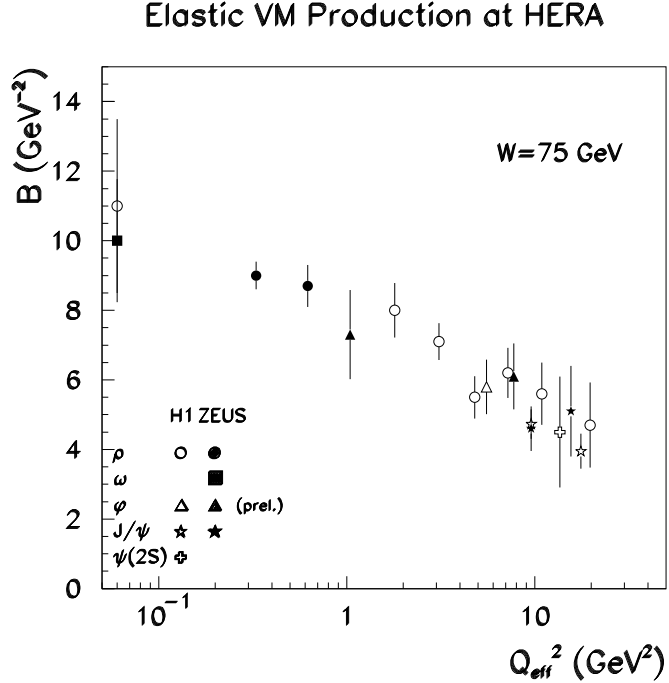,height=10cm,width=10cm}}
\caption{\sl Slope parameter $B(s)$ as a function of $Q_{eff}^2$;
$Q_{eff}^2= Q^2$ for the $\rho$ and $\omega$ meson production,
$Q_{eff}^2= Q^2+M_{\phi}^2$ and $Q_{eff}^2= Q^2+M_{J/\Psi}^2$ for 
the $\phi$ and $J/\Psi$ production.
}
\label{fig:bslope}
\end{figure*}

As we discussed after eq.(7), the slope $B(s)$ of the exponential 
fall-off of the observed $t$ distribution of the elastic cross section 
determines the size of the interaction region: $R_{int}^2 =2 B(s)$. 
In Fig.~\ref{fig:bslope} we 
first show the slope $B$, at a fixed energy, as a function of the
effective $Q^2$ scale, $Q^2_{eff}$, for various vector meson production 
processes~\cite{ref.vm}.  
The slope $B$ (and the size of the interaction region) shrinks with 
increasing virtuality $Q^2_{eff}$. 
In the photoproduction region, $Q^2_{eff}=0$, the observed 
$B$ values of $\rho$ and $\omega$ are of the order 10 GeV$^{-2}$, 
and they are similar to the 
values observed in  proton-proton scattering 
at high energies (with $B$ of the order $ 12-14$  GeV$^{-2}$).  
At higher $Q^2_{eff}$, the $B$ slopes at HERA become
considerably smaller, $B \approx 4-5$ GeV$^{-2}$: this is, approximately, 
half of the $B$ value observed in proton-proton scattering at low energies, 
i.e. it corresponds to the ``interaction size of a single proton''.
This suggests that
the transverse extension of the scattering system 
has the form shown in Fig.~\ref{fig:disc}b: the interaction region 
is determined by the size of the proton, whereas the photon and its
quark-antiquark pair 
(denoted by the black point) have radii much smaller than the proton. 
Its perturbative radiation cloud (small circles around the black dot) 
determines the growth  of the cross section with energy, and its size is 
smaller than the proton radius. The nonperturbative continuation of the radiation 
cloud (not shown), presumably, contains large-size wee partons, but they 
seem to be ``dormant'', in that they have little influence on the growth of 
the total cross section.
%Consequently, there seems to be not much space for the 
%development of a nonperturbative wee parton cloud around 
%the scattering system: the situation is quite different from hadron-hadron 
%scattering. 

Let us next see how the transverse shape of the 
scattering system evolves with energy. 
In eqs.(~\ref{eq:alphat})-(~\ref{eq:bs}) we have 
described, for hadron-hadron scattering, how the $t$-slope of $\alpha(t)$, 
the exponent of the energy dependence of the scattering amplitude,
leads to an energy dependent contribution to $B(s)$ and thus    
determines the growth of the scattering system in the transverse direction.
\begin{figure*}[htb]
\centerline{
\epsfig{file=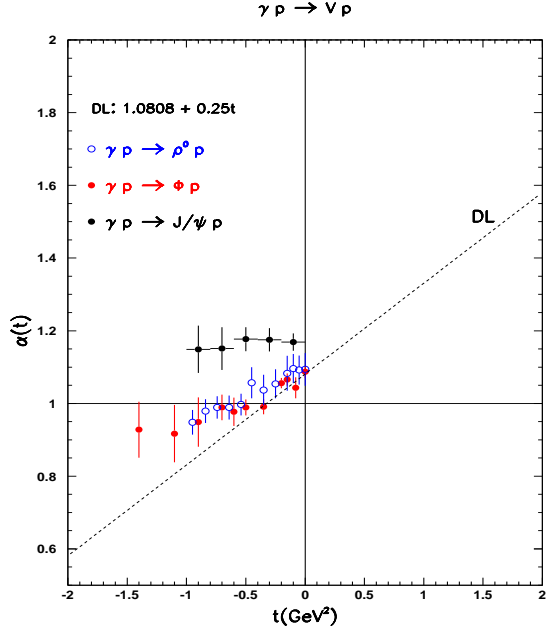,height=10cm,width=8cm}
}
\caption{\sl $\alpha (t)$ for diffractive $\rho$, $\phi$ 
and $J/\Psi$ photoproduction.}
\label{fig:zeus.a-vector}
\end{figure*}
Figure~\ref{fig:zeus.a-vector}
shows the measured $\alpha(t)$ dependence as a function of $t$ 
obtained from a 
combination ~\cite{ref.alphat} of the HERA photoproduction data and from low 
energy measurements from other experiments. For light vector 
particles ($\rho$, $\phi$) which have transverse radii of hadronic size the 
exponent $\alpha(t)$ is rather close to the hadronic Pomeron trajectory 
$\alpha_{\fP}(t)$ (denoted by ``DL''). 
In the case of $J/\Psi$ photoproduction 
which belongs to the class of small-size diffractive systems the
exponent $\alpha(t)$ shows a weaker $t$ dependence, i.e. $\alpha'$ is
significantly  smaller than $\alpha_{\fP}' = 0.25$ GeV$^{-2}$. 
%(further analysis of HERA data is in progress).
We view this smallness of $\alpha'$ for $J/\Psi$-production as the 
{\it second striking difference} 
in the elastic scattering of small-size and hadron-size projectiles:
not only the energy dependence of the scattering cross section at $t=0$, 
but also the $\alpha'$ parameter in the exponent exhibits different 
characteristics. Calculations in perturbative QCD lead to the expectation
that $\alpha'$ should be small (section 3.4), i.e. they are in qualitative 
agreement with the HERA measurement.

In order to obtain a geometric interpretation (in transverse direction) 
of $J/\Psi$ photoproduction at HERA let us first consider the situation 
(until now still hypothetical) of a system which has 
a much smaller size than the $J/\Psi$ system at $Q^2=0$. 
In this case perturbative QCD predicts a vanishing slope parameter $\alpha'$ 
which, by the arguments 
given in section 2, leads to the conclusion that the interaction radius 
does not increases with energy. This is illustrated in the    
transverse picture shown in Fig.~\ref{fig:disc}b: the proton size does not 
change with energy, and large-size wee partons whose cloud might 
grow with energy are ``not active''. Evolution with energy takes place 
mainly inside the 
radiation cloud around the $q\bar{q}$ pair: with increasing energy the 
number of radiated 
gluons grows, and the density inside the cloud around the black dot  
increases.  

The key question now is the change of the radiation cloud 
when $Q^2$ is taken to be smaller, i.e. the diffractive system becomes
larger. Since we do not yet have a satisfactory theoretical understanding,
only some first qualitative ideas can be formulated.
We start from large $Q^2$ values illustrated in Fig.~\ref{fig:disc}b, 
where we have marked only the perturbative part of the radiation cloud around 
the $q\bar{q}$ pair, which is responsible for the rise of the 
cross section. The extension of this region is approximately 
$\sim 1/ \sqrt{Q_0^2}$, which
is smaller than the proton radius. If we now move to smaller values of
$Q^2$ (larger $q\bar{q}$ pairs), the size of the gluons inside this
region will increase, and this will lead to a higher parton density. 
This favors the changes in the radiation cloud which we have mentioned 
at the end of section 3.1: additional gluon cascades are generated 
which damp the increase with energy. As a result, this region is no 
longer dominant over the large-size wee partons (not shown in Fig.
~\ref{fig:disc}b) which were ``dormant'' at large $Q^2$ and are now becoming
active. For very small $Q^2$ (and light quark systems), the wee partons 
have completely taken over, 
and the scattering system behaves like in hadron hadron scattering. 

Returning to the HERA data on the $\alpha'$-slope in $J/\Psi$ photoproduction,
it is tempting to interpret them as the onset of these  
transition phenomena. The measured $\alpha'$ value is small but, 
maybe, nonzero: the interaction size could already be slowly 
growing with energy. Compared to the situation illustrated in 
Fig.~\ref{fig:disc}b which belongs to the large-$Q^2$ region, 
this would mean that the wee partons have started developing their cloud 
at the surface of the proton. Presently measurements of the $\alpha'$-slopes
in various vector meson processes are in progress, and considerable
improvement in data precision is expected.

\subsection{Theoretical analysis of the $\gamma^* p$ processes}

In this section we discuss in somewhat more detail how, 
in the proton rest frame,  perturbative QCD 
calculations (which are now well-established) lead to the radiation picture 
which we have described so far. In the second 
part we will turn to current (and, naturally, less well-established) 
ideas on the transition 
from the small-distance to the large-distance region which are now being 
studied. 

Let us first return to the gluon-emissions illustrated in       
Fig.~\ref{fig:space-time}. As outlined before,  the 
creation of the $q\bar{q}$ pair takes place a long time before the
photon reaches the proton target. In order to estimate the lifetime of
a $\gamma^* \to q\bar{q} \to \gamma^*$ fluctuation we start from the 
four momenta of the proton and the photon in the proton rest frame: 
$$p^{\mu}=\left( m_p,0,0,0 \right)$$ 
and 
$$q^{\mu}=\left(Q^2/2xm_p,0,0,\sqrt{Q^2+(Q^2/2xm_p)^2}\right).$$ 
The incoming photon splits into the $q\bar{q}$ pair with momenta 
approximately equal to $z q^{\mu}$ and $(1-z) q^{\mu}$
(where $0< z <1$), and the energy difference between 
the photon and the $q\bar{q}$-state (which, according to the rules of 
old-fashioned noncovariant perturbation theory~\cite{OFPT}, 
consists of on-shell quarks) is $\approx x m_p$. 
Hence we estimate the lifetime as
\begin{eqnarray}
\tau = 1/m_p x
\end{eqnarray}
which in the low-$x$ part of the HERA region can be very large. In the same 
way one can show that 
the time interval between emission and absorption of the 
gluon with momentum  
$k_1$ is shorter than $\tau$, but longer than the corresponding time 
interval for the second gluon with momentum $k_2$. 
This explains the ``nested'' time structure of gluon emissions shown in   
Fig.~\ref{fig:space-time}.

The further development of the gluon emission processes 
can be deduced from the QCD Feynman diagrams that give the leading 
contribution in the 
high-energy limit of $\gamma^*p$ scattering; in a physical gauge 
(axial gauge), they have the ladder-like structure shown in 
Fig.~\ref{fig:space-time}, and non-ladder-like diagrams are suppressed. 
In order to obtain the space time
picture, we have to translate the covariant Feynman amplitudes into 
old-fashioned time-ordered perturbation theory~\cite{OFPT}:  
we write all possible time orderings and find that, in the limit $x\to 0$ 
($W^2 \to \infty$), the time ordering shown in Fig.~\ref{fig:space-time}
dominates. In these diagrams the momenta are conserved whereas the 
energies are not (this feature allows, for example, to estimate the 
lifetime (13) of photon fluctuation). The main result of this 
``old-fashioned perturbation theory'' is the time-ordering 
indicated in 
Fig.~\ref{fig:space-time}: the emission of the gluon with momentum $k_1$
happens after the creation of the $q\bar{q}$ pair, the emission of the
second gluon (with momentum $k_2$) is later than the first gluon etc. 
Taking further into account that quarks and gluons are traveling with the 
speed of light, one obtains an analogous ordering in the longitudinal spatial
direction ($z$-direction). As a result, the picture in 
Fig.~\ref{fig:space-time} describes the evolution both in time and in the
longitudinal $z$-direction.   

We now return to 4-dimensional momentum space and describe the 
kinematic region of the emitted gluons. The $q\bar q$ pair
emits first a gluon  with  four-momentum $k_1$. This gluon carries the 
fraction $z_1$ (with  $0< z_1 < 1$) of the photon longitudinal
momentum $q_l$, and the transverse component of its momentum, $k_{1t}$,
is on the average somewhat smaller that $\sqrt{ Q^2}$. In the next step 
this first gluon emits a second gluon with   
momentum $k_2$, which carries a fraction  $z_2$ of the photon longitudinal 
momentum  (with  $0< z_2 < z_1$) and has a transverse momentum, $k_{2t}$, 
which is on average somewhat smaller than $k_{1t}$. The probability to
emit a gluon in the $i$th step of  this cascade is found to be proportional 
to        
\begin{eqnarray}
\frac{3\alpha_s}{\pi} \int dy_i \int \frac{d^2 k_{it}}{k^2_{it}} 
\label{eq:prob}
\end{eqnarray}
where $\alpha_s$ denotes the strong coupling constant and $y_i$ the rapidity
of the emitted gluon, $y_i \approx \ln(z_i W^2/k_{it}^2)$.
In each step of this cascade the longitudinal and  transverse momenta
of the gluons decrease. 
The perturbative cascade process breaks down when the 
transverse momentum of the last gluon becomes too small. It is customary to 
assume for $Q_0^2$ a value of the 
order of a few GeV$^2$,
such that the transverse size of the last gluon with momentum 
$k^2_{nt} \approx Q^2_0$ is still substantially smaller than the proton.
This last gluon, which interacts with the proton in a nonperturbative way, 
carries the smallest fraction of the large photon momentum and, 
among the partons of the perturbative cascade, comes closest to the wee 
parton described in section 2. 
After this last gluon has finished its interaction with the proton 
all the emitted gluons are reabsorbed into the $q \bar q$ pair which at  
the time $\tau /2$ annihilates again into the virtual photon. 
With increasing energy the cascade of gluon emission becomes 
longer: more and more gluons are emitted.

Before we write down the full expression for the emission of the $n-1$ gluons
shown in Fig.~\ref{fig:space-time} we note that, from general
rules of quantum field theory, the 
radiation of a single field quantum of spin $J$ produces a total 
cross section proportional to $\left(W^2\right)^{2J-2}$. Since the gluon 
carries spin 1,  the radiation of the first gluon already produces a constant
total cross section, and each additional 
radiation will lead to a further logarithmic enhancement.   
The sum over $n$ of the Feynman diagrams 
shown in Fig.~\ref{fig:space-time} leads, in the limit of 
high $Q^2$ and low $x$, to the
following simplified ``radiation formula'':   
\begin{eqnarray}
\sigma_{tot}^{\gamma^*p}=
C \sum_n \left( \frac{3 \alpha_s }{\pi} \right)^n 
\int dy_1 \int \frac{dk_{1\;t}^2}{k_{1\;t}^2}...\int dy_{n-1} 
\int \frac{dk_{n-1\;t}^2}{k_{n-1\;t}^2}
\int dy_n \varphi(y_n,Q_0^2). 
\label{eq:sigtot}
\end{eqnarray}
$C$ denotes the $n$-independent normalization factor which is of no 
importance for our present discussion. The factor $\varphi(y_n,Q_0^2)$ 
describes
the $y_n$ dependence of the interaction of the last gluon with the proton, 
resulting from the wee partons inside the proton which we cannot 
calculate. The emission of gluons is ordered in rapidities since the 
$y_i$ values are   
determined by the fractions $z_i$ of the photon longitudinal momenta
carried away by the gluons, and the $z_i$ are ordered. 
This defines the integration limits in $y$:
\begin{eqnarray}
0<y_n<...<y_1<\ln(1/x). 
\end{eqnarray}
The upper limit is given by the largest possible value of rapidity, 
$\ln (W^2/Q^2) \approx \ln (1/x)$.
In order to illustrate the most important feature of the radiation 
formula, the rise at small $x$, we first present a very simplified argument. 
We will assume 
$\varphi(y_n,Q_0^2) \equiv 1$, neglect the decrease of 
the transverse momenta,  
and we make the assumption that the integrals over the 
transverse momenta are independent of each other and range from the low 
momentum scale $Q_0^2$ up to $Q^2$. 
One easily sees that the radiation formula 
(at fixed $Q^2$) then leads to the simple power behavior    
\begin{eqnarray}
\sigma_{tot}^{\gamma^*p} \sim \left( 1/x \right)^{\lambda} 
\label{eq:xtol2}
\end{eqnarray}
with $\lambda=b \frac{3 \alpha_s }{\pi}$, 
$b=\int_{Q_0^2}^{Q^2} \frac{dk_t^2}{k_t^2}$ 
(the ordering of rapidities gives the factor  $1/(n-1)!$ which allows to 
exponentiate in eq.~\ref{eq:sigtot}). The value of 
$\lambda$, because of the $k_t$ integration limits, is expected  
to increase with $Q^2$. The relation (~\ref{eq:xtol2}) is
equivalent  to the relation 
$\sigma_{tot}^{\gamma^*p} \sim \left( W^2 \right)^{\lambda}$ 
%used to extract (at fixed $Q^2$) the $\lambda_{tot}$ values from data 
since in the low $x$ region $1/x = W^2/Q^2$. 

In order to be more accurate we have to go one step further and discuss, 
in the radiation formula eq.(\ref{eq:sigtot}), the range of 
integration of the transverse momenta which depends upon details of the 
kinematic limit. So far we have not been very 
explicit about this aspect: in order to have a small-size
incoming photon we have to consider large $Q^2$ values. On the other hand,
we want to have large energies $W$; this 
leads us to the combined limit of large $Q^2$ and small $x$. Depending now 
upon the precise order of these two limits, two different restrictions on the
transverse momentum integration are obtained. In the classical deep inelastic 
limit (``Bjorken limit'') one first takes $Q^2$ large, then $x$ small. In this
limit one finds the following ordering of the transverse momenta: 
\begin{equation}
k_{nt}^2 < k_{n-1t}^2 < ... < k_{1t}^2 < Q^2.
\end{equation} 
Alternatively,  one considers first $x$ to be small, then $Q^2$ 
to be large. In this case the transverse momenta are not ordered. 
For simplicity we
choose:
\begin{equation}
Q_0^2 < k_{it}^2 < Q^2, \,\,\,\,i=1,...,n
\end{equation}
(in this case we also have to modify the emission formula
(15) by a kernel $K(k_i, k_{i+1})$). If these restrictions are taken into 
account we find for the radiation formula:
\begin{equation}
\sigma_{tot}^{\gamma^*p}\sim \left\{ 
\begin{array}{ll} 
\exp{\sqrt{\frac{4\alpha_s N_c}{\pi}\ln 1/x \ln Q^2/Q_0^2}} &(DGLAP)\\ 
    (1/x)^{\lambda} & (BFKL) 
\end{array}
\right\},
\end{equation}
where the former case (called DGLAP ~\cite{DGLAP}) belongs to the 
Bjorken limit, the latter one (called BFKL ~\cite{BFKL}) to the other ordering
of the two limits. Both prescriptions lead to a strong growth of the cross
section in $1/x$. At large $Q^2$ values, the DGLAP case applies. But at lower
$Q^2$ and very small $x$ values the HERA 
kinematic region, most likely, lies on the interface of the two limits. 
In both cases, the simple formula (~\ref{eq:sigtot}) leads to a 
growth of the total $\gamma^*p$ cross section and explains the observed 
rise.

A closer look at the transverse momenta also allows  the 
spatial extension in the transverse direction to be understood. 
Transverse momenta are conjugate to transverse distances: the fact that the 
transverse momenta of the emitted gluons start by being large, then 
decrease from one emission to the 
next, implies that the mean transverse distance between two subsequent steps 
of emission grows. Therefore the radiation cloud of the incoming photon, 
which started off as a small $q\bar{q}$ pair, grows in the transverse 
direction. Since the mean separation between steps of gluon emissions grows,
this growth is faster than, e.g., in normal diffusion. When the transverse 
momenta of the emitted gluons become small, perturbative QCD ceases to be 
applicable: the scale $Q_0^2$ at which this happens corresponds to a spatial 
transverse extension which has to be smaller than the proton radius. 
For larger transverse distances, most likely, the evolution of the radiation
cloud continues: in eq.(~\ref{eq:sigtot}) this long-distance part of the 
radiation cloud is parameterized by the function $\varphi(y_n,Q_0^2)$, and it 
has its own (weak) energy dependence which has to be taken from 
experimental data. This discussion leads us to the picture illustrated in 
Fig.~\ref{fig:disc}b: the incoming photon
(denoted by the black dot) is much smaller than the proton. It is surrounded 
by its perturbative radiation cloud (denoted by small circles) which 
produces the strong increase of the cross section with energy. Its long 
range, nonperturbative part (not shown) reaches the full proton size: its 
influence on the energy dependence of the total cross section is much weaker 
than that of the perturbative part of the cloud.  

To finish our summary of perturbative QCD calculations we have to address 
the question of the $t$-dependence of the scattering amplitude. 
The main result is that, for small-size photons and small momentum 
transfer, the energy growth of the scattering amplitude varies 
only little with $t$. This can easily be made plausible by 
generalizing the QCD radiation formula to nonzero momentum 
transfer. Let us consider a small momentum transfer (e.g. $|t|<0.5$ 
GeV$^2$). Starting from Fig.~\ref{fig:space-time} we 
give the incoming and outgoing photons small transverse momenta,
$-\vec{k}/2$ and $\vec{k}/2$, resp., with $t=-\vec{k}^2$. Correspondingly, 
the gluons on the l.h.s. 
(with four momenta $k_1,...k_n$) carry the additional 
transverse momentum      
$-\vec{k}/2$. We have argued, in order to obtain the radiation formula, the 
transverse momenta have to be large (up to the order of $\sqrt{Q^2}$): 
this means that, to a good approximation,
the small momentum transfer $\vec{k}/2$ can be neglected, and the result for the energy dependence remains practically 
unchanged when going from $t=0$ to $t\neq 0$. The same argument can 
(approximately) be used 
for diffractive $J/\Psi$-production, and it explains the observed small 
$t$-slope $\alpha'$ of the trajectory $\alpha(t)$ shown in 
Fig.~\ref{fig:zeus.a-vector}.
Following our previous argument for the geometric interpretation of 
$\alpha(t)$, we therefore conclude that the transverse extension of the 
scattering system shown in Fig.~\ref{fig:disc}b does not increase 
with energy: 
the density of the gluons around the $q\bar{q}$ pair increases, but
the transverse radii of the pertubative cloud and of the proton 
remain constant.
   
This completes our discussion of the small-size scattering process
(photon with large $Q^2$ or vector meson with heavy quarks). 
We now turn to the transition to large distances. 
This is where theoretical interest is now focused, and
for which intensive comparisons with experimental data are being made.  
We have already indicated what kind of ideas are being pursued:
at low $Q^2$, before the perturbative radiation cloud 
(consisting of a single gluon cascade) develops into the nonperturbative 
cloud of wee partons, we first expect to see the beginning of 
$2$, $3$, ... cascades which lead to 
the production of a large number of gluons.
In other words, we expect that before the long-distance physics 
of QCD confinement begins we may see an intermediate 
region which can be addressed still using quark and gluonic
degrees of freedom.
One of the main characteristic features is the high density of gluons which
generates a strong gluon field. If true, this is a rather remarkable 
state: on the one hand, it is still dominated by rather short 
distances where typical confinement effects (e.g. chiral symmetry breaking 
and the formation of pion pairs which are an essential part of the 
nonperturbative wee partons) can be disregarded. On the other hand, the high 
density of gluons creates a strong field and leads to sizable cross 
sections. This expectation is based upon extensive studies of higher order
QCD corrections to the single-cascade picture shown in Fig.~\ref{fig:space-time}
~\cite{Sat}.
In order to give an intuitive motivation for the appearance of this novel 
high-density state in QCD,  
let us consider a situation where, at fixed large $Q^2$, $x$ becomes 
extremely small, much smaller than at HERA. Since the increase
of the total $\gamma^*p$ cross section with energy, as predicted by the single
gluon cascade, cannot continue forever, there must be some corrections which
slow down the rise. The one-cascade picture can be valid only 
if the system of gluons is sufficiently dilute. When at very small $x$ the 
gluon density of the single cascade becomes too high, interactions 
with other cascades (which are always present but, in the standard case of not 
so small $x$, are simply not noticed by the proton) will become important. 
An example is the formation of two separate gluon cascades: 
after the emission of the 
first gluon (in Fig.~\ref{fig:space-time} with momentum $k_1$) which 
initiates the first cascade of gluon emissions, 
the $q\bar{q}$-pair radiates a second gluon which begins its own cascade.    
This formation of two branches of gluon radiation becomes relevant
when the offsprings of both cascades have a chance to simultaneously 
find partners inside the proton. Such pairwise (coherent) interactions
of both clouds lead to negative interference effects and damp the energy  
growth of the cross section. As $x$ decreases further, more and more 
cascades are formed; the rise in $1/x$ of the gluon density 
becomes weaker, and the gluon density could even become flat in $1/x$, 
i.e. it reaches ``saturation''. This interplay of a large number of cascades  
which leads to saturation effects has to be viewed as a precursor of 
confinement: once saturation has eliminated the dominance of the small-size
part of the radiation cloud, the large size (of the order of the 
proton radius) wee partons and long-range properties of the 
QCD vacuum become important, and quarks and 
gluons are no longer the suitable degrees of freedom to describe
the radiation cloud. 
 
It is very plausible to expect that the energy value at which these 
saturation effects set in will go up as $Q^2$ increases: the smaller the 
transverse size of the $q\bar{q}$ pair and of the radiated gluons, the longer
the cascade will grow before the state of high density is reached 
and the formation of additional gluon cascades becomes relevant. 
Conversely, at low $Q^2$ saturation sets in earlier: this suggests that 
the observed damping of the growth of the cross section at low $Q^2$ 
(or at least its beginning) may be due to this same saturation 
mechanism ~\cite{Almueller}. The success 
of a simple model,  described below,  supports this conjecture. 
 
As we have said before, these questions are currently under investigation. 
Most recently, attempts have been made to find a phenomenological 
{\it interpolation formula} which, by comparing to HERA data,
could test these ideas. For such an attempt it is useful to have a 
formalism which is general 
enough to provide a bridge from small to large transverse distances and to 
apply to both the total cross section and to the diffractive cross 
sections. In the proton rest frame the total cross section, to good 
approximation, can be written in the following form:
\begin{eqnarray}
\sigma_{t,l}^{\gamma^* p}(x,Q^2) = 
\int d^2 \vec{r} \int dz \, \psi(Q^2,z,\vec{r})_{t,l}^*
\, \sigma_{q\bar q p}(x,\vec{r}) \psi(Q^2,z,\vec{r})_{t,l},
\label{eq:satur1}
\end{eqnarray}
\noindent 
where $\psi(Q^2,z,\vec{r})_{t,l}$ denotes
the transversely $(t)$ and longitudinally $(l)$  polarized 
 photon wave-function, $\sigma_{q\bar q p}(x,\vec{r})$ 
the dipole cross section of the interaction of the   
$q\overline{q}$ pair with the proton, $z$  the momentum fraction of the 
photon carried by the quark, $0<z<1$, 
and $\vec{r}$ the transverse size of the quark-antiquark pair.
The wave functions are solely determined by the coupling
of the photon to the quark pair, and they are well known in QED.
Eq.(~\ref{eq:satur1}) corresponds to the
physical picture that we have drawn: the incoming virtual photon creates
a quark-antiquark pair (expressed in terms of the photon wave function
$\psi$), the $q\bar{q}$ pair
interacts with the proton (denoted by $\sigma_{q\bar{q}p}$), and
finally annihilates into the outgoing virtual photon ($\psi^*$). 
All interesting dynamics (gluon radiation etc.) is encoded in the 
$\vec{r}$ and $x$ dependence of $\sigma_{q\bar{q}p}$. For small
$\vec{r}$ we have the gluon radiation which we can calculate in perturbation 
theory, and we find that the cross section is approximately proportional to 
$\vec{r}^2$. At medium $\vec{r}$ values we expect to see the beginning of  
``saturation'' that we have sketched before, and for large
$\vec{r}$ the dipole cross section 
has to become rather flat in order to agree 
with the observed weak energy growth of the hadronic cross section at low 
$Q^2$. In eq.(~\ref{eq:satur1}) the main contribution of the $\vec{r}$ 
integral comes from distances $|\vec{r}|$ around (or slightly larger than) 
$2/\sqrt{Q^2}$.
 
\begin{figure*}[htb]
\centerline{
\epsfig{file=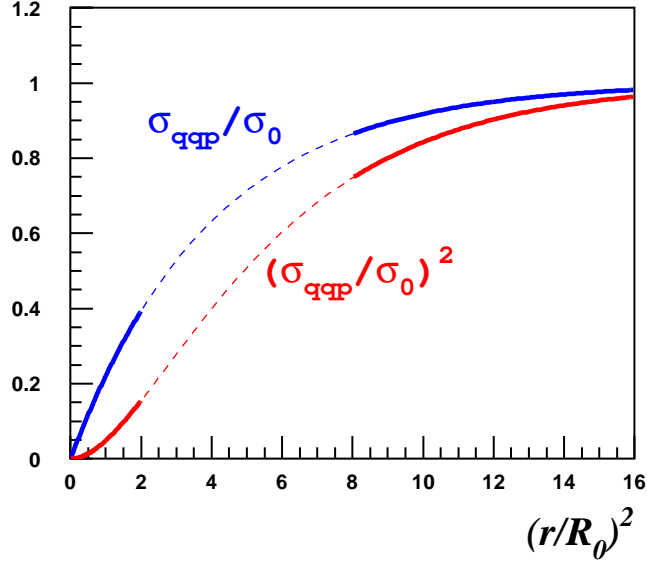,height=9cm,width=12cm}}
\caption{\sl The dipole cross section $\sigma_{q\bar q p}$ 
of ~\cite{GBW} as a function of the transverse extension of the 
quark-antiquark pair. 
The lower graph shows the square of this cross section which is relevant
for the inclusive diffractive process }
\label{fig:fgrassqq}
\end{figure*} 

The dipole cross section $\sigma_{q\bar{q}p}$ can also be used  to describe 
diffractive processes, in particular vector meson production or the 
inclusive diffractive cross section 
(in the forward direction $t=0$). For example, the integrated cross section of diffractive
$q\bar{q}$ production can be expressed in terms of the dipole cross section:
\begin{eqnarray}
\frac{d\sigma^{\gamma^* p}_{Diff}}{dt} \left|_{t=0} \right.  = \frac{1}{16\pi}
 \int d^2 \vec{r} \int dz \, 
 \psi (Q^2,z,\vec{r})^* \sigma^2_{q\bar qp}(x,r^2)\,
\psi (Q^2,z,\vec{r}) 
\label{eq:saturdiff}
\end{eqnarray}
This expression does not depend any more on any wave-function of the 
diffractive final
state: the $q\bar{q}$ pair is no longer forced into a final state of a 
given size (as it was, for example, the case for vector particles),
but can choose rather freely its preferred size. A large fraction of the 
events, in fact, selects configurations in which the quarks have rather large 
transverse extensions (``aligned jets''). In contrast to the total cross 
section formula, eq.(~\ref{eq:satur1}), which is linear in the dipole cross 
section, the inclusive diffractive cross section, 
eq.(~\ref{eq:saturdiff}), contains the square. As we have mentioned before, 
in any dipole model at small $\vec{r}$ the cross section is approximately 
proportional to $\vec{r}^2$.
Therefore the contribution of final states with small $\vec{r}$ is suppressed
in comparison with the total cross section (for an example see 
Fig.~\ref{fig:fgrassqq}:
$(\sigma_{q\bar{q}p}/\sigma_0)^2 \ll \sigma_{q\bar{q}p}/\sigma_0$ at 
$r \ll 2 R_0$). This is in agreement with 
the observed large transverse size of the inclusive diffractive process 
($B\approx 7$ GeV$^{-2}$ at $Q^2>5$ GeV$^2$ ~\cite{LPS}).    

We conclude our discussion with a recent model ~\cite{GBW} for the dipole 
cross section which has been built on the idea of saturation and which has 
turned out to provide a very successful description of the HERA data in the 
low $Q^2$ region. The ansatz for the dipole cross section is: 
\begin{eqnarray}
\sigma_{q\bar q p}(x,\vec{r}) = 
\sigma_0 \, (1 - \exp [-\frac{r^2}{4R^2_0}]),
\label{eq:satur2}
\end{eqnarray}
where $R_0$ denotes the ($x$-dependent) dipole size where saturation sets 
in (''saturation radius'': $R_0^2=R^2_0(x) = \frac{1}{Q^2_0}  
\left( \frac{x}{ x_0} \right)^{\lambda_{GBW}}$ with $Q_0=1$ GeV).
Figure~\ref{fig:fgrassqq} shows the graphic representation of 
$\sigma_{q\bar qp}$. 
The parameters of the model, $\sigma_0$, $x_0$, and $\lambda_{GBW}$, were
determined from the fit to the total cross section data: $\sigma_0=23mb$,
$\lambda_{GBW}=0.29$, and $x_0 = 3 \cdot 10^{-4}$.
The dashed curves in Fig.~\ref{fig:sigtotw2} 
and  Fig.~\ref{fig:lam} show that
in the low-$Q^2$ region the energy dependence of the total cross section is 
very well described by the model. The model has also been used for
diffraction: when the formula for the dipole cross section is inserted into 
(eq.~\ref{eq:saturdiff}), it leads to a 
good description of the diffractive cross section for $q\bar{q}$ production.
In particular, as a result 
of a subtle interplay between the photon wave function and the dipole cross 
section, it succeeds in reproducing the energy dependence shown in 
Fig.~\ref{fig:zeus.rdifftot} (note that the results of 
the model shown in this
figure are genuine predictions, since all the parameters of the model
have already been fixed  by the measurements of the total 
$\gamma^{*}p$ cross section). 

Although the dipole formula in eq.~\ref{eq:satur2} should primarily be 
considered as a phenomenological
parameterization of the transition region between small and large
sizes of the $q\bar{q}$-pair, it nevertheless illustrates some features 
of the peculiar ``high density'' region. First, for small dipole sizes $r$    
one sees the dipole behavior $\sim r^2$. Next, if we increase the energy 
(i.e. we go to very small $x$, beyond the HERA region), $R_0$ becomes rather 
small, and the exponent in the dipole formula is of 
order unity even for small dipole sizes $\vec{r}^2 \approx 4 R_0^2$: 
since $\sigma_0=23$mb is a large 
cross section (comparable to the $\pi p$ total cross section), 
we arrive at a large dipole cross section already in the region of small 
distances, $\vec{r}^2 \approx 4 R_0^2 \ll R_p^2$, where typical long distance 
confinement effects are not yet expected to be important. 
This large cross section is rather due to 
the high gluon density which produces a strong gluon field. A more detailed 
analysis, in fact, allows  the dipole cross section formula, 
eq.(~\ref{eq:satur2}), in this intermediate region to be interpreted as 
a sum of multi-gluon cascade contributions, 
very much in the way we have outlined before.
For dipole sizes larger than $R_0$, we can disregard
the exponential term: this is then the long-distance region 
where confinement 
holds. The success of this simple model indicates
that the gross features of the physical picture which underlies  
this dipole model must be correct. The next step could be a generalization
to nonzero momentum transfer: this addresses the generation of the slope 
parameter $\alpha'$. We have argued before (and seen in the data) 
that in the perturbative region 
$\alpha'$ is small. But we also know that at large distances 
it has to  
approach the hadronic value $\alpha_{\fP}' =0.25$ GeV$^{-2}$.
Understanding the origin of this confinement parameter presents the most 
challenging task.
 
Summarizing this ``theoretical section'', we have shown that QCD 
calculations are in qualitative 
(partially also quantitative) agreement with the small distance parts of the 
observed $\gamma^*p$ scattering. HERA data at low $Q^2$, which probe 
large distances, are fully consistent with  previously observed 
hadron-hadron scattering. We therefore have to develop a formulation which
allows an interpolation between the perturbative QCD calculations at short 
distances and the nonperturbative features of hadron-hadron scattering 
at large transverse distances.
The suitable physical picture is {\it radiation}: at small distances 
QCD calculations, 
when interpreted in the proton rest frame, lead to the simple 
picture of a {\it radiation cloud of the incoming photon}, and at large 
distances we know the general features of 
the wee partons which again fit into the framework of radiation. 
For the transition from the perturbative QCD region to the confinement 
region we have precise data which allow a test of any theoretical hypothesis.
On the theoretical side there is the well motivated expectation that 
between the regions of perturbative QCD and confinement dynamics there 
exists a novel region of ``high density partons''. The simple saturation 
model ~\cite{GBW} which is based upon these ideas has turned out to be 
very successful and thus provides phenomenological support for the 
underlying dynamical mechanism.

\begin{figure}[ht]
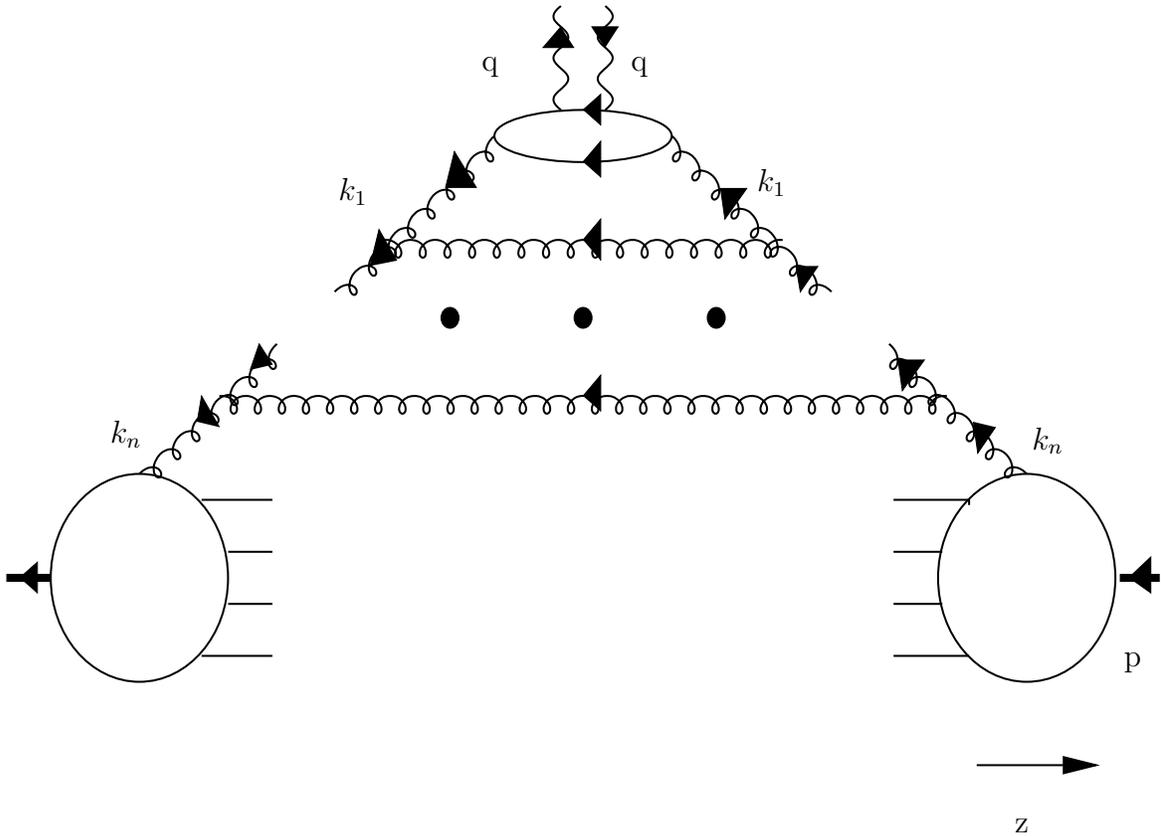

\vspace{2.cm}
\begin{center}
\input kasten2.pstex_t
\end{center}
\caption{\sl Space-Time diagram of the elastic $\gamma^* p$ scattering
process in the deep inelastic (Bjorken) frame.}
\label{fig:space-timeb}
\end{figure}  
\section{Deep Inelastic Structure Functions}
How is this picture of a ``radiation cloud of the incoming photon'' 
related to the concept of the ``deep inelastic proton 
structure function''? The precise measurement of these structure functions 
and of parton densities in 
a wide kinematic region is one of the important goals at HERA.
Parton densities present probability densities for finding partons 
(quarks or gluons) with specified momentum fractions and transverse extensions
inside the proton. In order to have an intuitive picture of ``partons inside
the proton'' we have to go into another Lorentz-frame, the 
so called Bjorken frame, 
in which the proton has a much larger longitudinal momentum than 
the photon. In this frame the four momenta have the form:
\begin{eqnarray}
p^{\mu}=\left( \sqrt{p^2+m_p^2}, 0,0 -p \right),
\end{eqnarray}
(with a very large $p$) and
\begin{eqnarray}
q^{\mu}=\left( Q^2/2xp,\vec{q_t},0\right ),
\end{eqnarray}
with $\vec{q}_t^2\approx Q^2$. 
The space-time picture of the process is obtained in the same way as in  
in the proton rest frame: the dominant QCD Feynman diagrams are
translated into time-ordered (old-fashioned) perturbation theory for
which it can be shown~\cite{OFPT} that only the time-ordered diagrams 
of  Fig.~\ref{fig:space-timeb}  contribute to the high energy limit.
Therefore now it is the proton (moving in the negative $z$-direction) 
which, long before it interacts with the photon, starts to 
build up its radiation cloud. Since the transverse extension of the
proton is large, the birth of this cloud cannot be described  in terms of 
perturbative quarks and gluons.
Only at a later stage of the formation process, when perturbative partons 
with sufficiently small transverse extensions have been created,  can  
this language start to be used. 
This ``final'' perturbative part of the radiation 
process is illustrated in Fig.~\ref{fig:space-timeb}.
Among these partons the incoming photon has to find its
partner with longitudinal momentum  equal to the fraction $x$ of the 
proton momentum, and with the large transverse momentum, 
$Q_0^2 \ll k^2_t \ll Q^2$,
The lifetime of the $q\bar q$ fluctuation which interacts with the
photon can be estimated as   $\tau_{q\bar{q}} \approx xp/<k_t^2>$. It is
considerably longer than the time between the absorption and emission of the
virtual photon, $\tau_{\gamma} =1/q_0 \approx 2xp/Q^2$: 
the parton cloud remains "frozen" during the interaction and 
can be precisely scanned by the virtual photon. 

The structure functions and parton densities obtained from the HERA data 
extend into kinematic 
regions which had not been measured before, in particular the region of very
small $x$ (i.e. large energy of the $\gamma^*p$-subprocess) and low
$Q^2$. The results which we have discussed in section 3 can easily be 
translated into the deep inelastic picture. Using the relation
\begin{equation} 
\sigma_{tot}^{\gamma^*p}= \frac{4\pi^2 \alpha_{em}}{Q^2}F_2(x,Q^2), 
\end{equation}
the observed strong growth with the energy of the total 
$\gamma^*p$ cross section is re-phrased
as an increase of the deep inelastic structure function $F_2$ at small $x$. 
One of the
important questions which still need to be answered is the limit of 
the applicability of the concept of parton densities: for small $Q^2$ 
(where the photon becomes hadron-like) the notion of partons and 
QCD perturbation theory breaks down, 
and we still have to determine more precisely how far down in $Q^2$ and $x$ 
the standard QCD analysis of deep inelastic scattering works.     

A comparison of this ``deep inelastic'' interpretation in the Bjorken frame 
with our previous discussion based upon the 
proton rest frame illustrates a very interesting feature. 
In the ``deep inelastic picture'' the radiation cloud results from the 
break up of the incoming proton and can therefore be named the
``radiation cloud of the proton''. In contrast, in the proton rest frame 
we have been discussing the ``radiation cloud of the photon''. 
When comparing the two pictures, one finds that it is mainly the time 
variable which plays different roles. On the other hand, we know that the
two pictures are completely equivalent: 
they are both derived from the same QCD calculations 
at small $x$ (or large energies $W$), by first choosing a frame
and then translating the Feynman diagrams into space time variables.
The cross section formulae are, obviously,  frame-independent.
The equivalence of the two pictures implies that the radiation cloud
belongs to both the proton and the photon simultaneously. This    
supports Feynman's conjecture that in a hadron-hadron collision the 
nonperturbative wee partons belong to both projectiles 
and are a fundamental part of the underlying strong forces.

For our discussion we found it more suitable to use the proton rest frame
 with the ``radiation cloud of the photon''.   
In order to follow the transition from radiation at short transverse 
distances to large transverse distances we 
found it instructive to chose that picture where the beginning of the
radiation cloud can be analyzed within perturbative QCD.
Later on, when the evolution of the radiation cloud 
enters the nonperturbative region, we have some knowledge based upon the 
wee parton analysis of high-energy hadron-hadron scattering. 

\section{Summary and Outlook}

Measurements at HERA have opened a new route to understand 
the confinement problem by the investigation of processes in which 
virtual photons with variable transverse sizes 
interact with  protons at high energies.
For photons with small transverse sizes, 
the measured cross sections exhibit a considerably faster growth with energy 
than in hadron-hadron scattering; this feature can be explained within
perturbative QCD. HERA data on diffraction show another important property: 
a scattering system consisting of a small-size vector particle and the proton 
has a transverse extension which is not larger than a single proton, 
and, in contrast to hadron-hadron scattering, it does not expand as energy 
increases. This observation is also in agreement with 
perturbative QCD expectations. The picture appropriate to understand these 
scattering processes is that of radiation: in the proton rest frame 
the perturbative QCD calculations can be formulated in terms of a 
radiation cloud of gluons and quarks which are emitted
within a femto-tube around the incoming virtual photon. 
For the first time in 
high-energy physics experiments, it is possible to understand the high 
energy behavior of cross sections which are mediated
by the strong forces.

The transverse size of the incoming photon can be varied continuously.
For photons with large transverse extensions, the cross sections
show the same characteristics as observed in hadronic processes.
The energy dependence of the cross sections in hadronic processes 
has long been known  to have universal properties 
( attributed to the
``Pomeron''). Attempts to understand the dynamics of this branch of 
strong interaction physics led, long before HERA, to an intuitive 
space-time picture of hadron-hadron scattering in which the radiation 
of wee partons plays an essential role. This picture is strongly supported
by the perturbative QCD cloud at short distances. 
A central task posed by HERA measurements is the need to understand 
the transition between QCD radiation at short distances and the 
nonperturbative wee parton dynamics at large transverse distances.
The HERA measurements and their theoretical interpretation 
show that  high-energy scattering can be 
viewed as  QCD radiation at a variety of transverse distances.
At both ends of the distance scale the radiation processes are, to a large 
extend, known. The small-size 
region is determined by QCD perturbation theory, the large-size 
domain by  universal hadronic properties. This opens an 
attractive possibility of interpolation
between these two extremes. First attempts in this direction have already 
been made, and, in particular, a simple model has been proposed which 
successfully describes 
the HERA data in the transition from perturbative QCD to nonperturbative 
strong interaction physics. HERA experiments have the potential further to 
increase the kinematically accessible measurement regions and to improve 
their measurement accuracy with increasing  luminosity. 
This will considerably 
increase the amount of informations available for this challenging  
transition region.

The novel approach to the confinement problem has raised the question
of understanding QCD radiation at large transverse distances. It is very
tempting to recall another moment in the history of physics where 
the investigation of a ``radiation'' problem has led to unexpected and 
far-reaching results: the black body radiation in electromagnetism.
In 1859 Kirchhoff had established that the energy distribution of black body
radiation is universal, i.e. independent of the material of the black body.
Shortly before the turn of the century Max Planck tried to find a theoretical
derivation of a formula for the energy distribution $J$. In a first step he
succeeded to interpolate between two known limiting cases (the Raleigh-Jeans
formula for small frequencies or long wave lengths, and Wien's law for
high frequencies or short wave lengths). When trying to ``derive'' his
interpolating formula he was forced to introduce the concept of ``energy 
quantization'',  as the only way to explain the high frequency limit. 
This was the beginning 
of quantum theory, but it took another quarter of a century before 
the theory for the energy spectra of atoms was formulated, and then another 
few decades before Feynman and Schwinger presented the modern form of Quantum
Electrodynamics which allows the calculation of  scattering and radiation 
processes involving charged particles. 
It is this framework which nowadays also allows 
computation of  the radiation of quarks and gluons in QCD at small distances.
Now a new step in understanding radiation is needed: at large
distances the QCD coupling becomes large, and confinement starts to
radically change  the radiation pattern.
It seems natural to first define quantities which can be measured both 
at small and at large distances (analogous to the energy distribution $J$ 
of the black body radiation). Presently  the study of the 
quark-antiquark dipole cross section, $\sigma_{q\bar{q}p}$, 
is attracting much interest and future theoretical work will concentrate 
on deriving a theoretical formula for this quantity.

There is no doubt that HERA has 
opened a new door into the complicated dynamics
of strong interaction forces. This article has attempted to provide 
a ``Bestandsaufnahme''. 
Understanding the dynamics of strong 
interactions remains one of the central tasks in particle physics. Also the 
search for fundamental theories beyond the Standard Model will find it 
necessary to have a better understanding of QCD dynamics: the development of 
new concepts (e.g. string theories) will need 
nonperturbative methods, for which QCD offers a challenging testing ground.
\section{Acknowledgements}
When writing this article we have profited from helpful  discussions 
with many colleagues. Our special thanks go to A. Caldwell, B. Foster, 
K. Golec-Biernat, T. Haas, D.Haidt, R. Klanner, E. Levin, E. Lohrmann, A. Mueller, 
S. Schlenstedt, and G. Wolf.

\end{document}